\documentclass[usenatbib]{mnras}

\usepackage{newtxtext,newtxmath}

\usepackage[T1]{fontenc}

\DeclareRobustCommand{\VAN}[3]{#2}
\let\VANthebibliography\thebibliography
\def\thebibliography{\DeclareRobustCommand{\VAN}[3]{##3}\VANthebibliography}

\usepackage{graphicx}	
\usepackage{amsmath}	
\usepackage{booktabs}
\usepackage{supertabular,booktabs}
\usepackage[tableposition=above]{caption}
\usepackage{longtable}
\usepackage{threeparttable}
\usepackage{dcolumn}
\usepackage{threeparttablex}
\usepackage{soul,xcolor}
\usepackage{xcolor}

\usepackage{pifont}
\newcommand\ddfrac[2]{\frac{\displaystyle #1}{\displaystyle #2}}
\newcommand{\RNum}[1]{\uppercase\expandafter{\romannumeral #1\relax}}
\setstcolor{red}

\title[The Galactic Isotopic Decomposition for the Sculptor dSph]{The Galactic Isotopic Decomposition for the Sculptor Dwarf Spheroidal Galaxy}

\author[Pandey $\&$ West]{
Kanishk Pandey,$^{1}$\thanks{E-mail: pandeyk@carleton.edu}
Christopher West$^{1, 2}$
\\
$^{1}$Department of Physics and Astronomy, Carleton College, Northfield, MN, 55057, USA\\
$^{2}$Joint Institute for Nuclear Astrophysics—Center for the Evolution of the Elements, USA
}

\date{Accepted XXX. Received YYY; in original form ZZZ}

\pubyear{2021}

\begin{document}
\label{firstpage}
\pagerange{\pageref{firstpage}--\pageref{lastpage}}
\maketitle

\begin{abstract}
Stellar evolution models require initial isotopic abundance sets as input, but these abundances are incomplete outside the solar neighborhood, are challenging to infer from elemental observations, and are galaxy specific. Compositions different from the Milky Way (MW) have distinct chemical histories and are important to explore. We present an isotopic history model for the Sculptor dwarf spheroidal galaxy (dSph) based on astrophysical processes, using a complementary approach to GCE models, which can estimate isotopic abundances for future nucleosynthesis studies. We approximated the isotopic composition of Sculptor’s late stage evolution using the OMEGA chemical evolution code and used Big Bang Nucleosynthesis (BBN) predictions as the other boundary condition. Isotopic abundances were scaled from late stage evolution to BBN values according to the astrophysical processes responsible for their production. The isotopic abundances were summed into elemental abundances and fit to observational Sculptor abundance data to tune the free parameters. The completed model gives the average isotopic history of Sculptor for massive star, Type Ia SNe, main $s$-process peak, and $r$-process contributions. We find that Type Ia SNe contribute $\approx$ 86 per cent to the late stage evolution Fe abundance, which agrees with other dSph chemical evolution studies, and is greater than typical MW values of $\approx$ 70 per cent found using a similar process. The model also finds that neutron star mergers contribute $\approx$ 30 per cent to the late stage evolution Eu abundance, suggesting that CCSNe may be the dominant $r$-process progenitors in dSphs.
\end{abstract}




\section{Introduction}
\label{Introduction}
Stellar simulations require an initial isotopic composition, and this input is crucial for obtaining the correct nucleosynthesis outputs. When modelling compositions similar to the sun, this is aided by a robust solar isotopic abundance composition. Many studies continue to refine this composition (e.g., \citealp{lodders2003, lodders2010, Lodders2020, asplund2005, asplund2009, grevesse2005, grevesse2007}). However, isotopic abundances are more challenging to determine for environments outside the solar neighborhood due to the lack of isotopic data, and such abundances cannot be inferred from elemental measurements. 

Galactic chemical evolution (GCE) models are one possible method to find an initial isotopic composition input. GCE models typically integrate stellar yields across time-steps to build abundances from big bang nucleosynthesis (BBN) to a final (e.g., solar) composition. In doing so, these models predict average isotopic abundances which can, in part, facilitate our understandings of stellar nucleosynthesis \citep{prantzos2008}. GCE models must estimate quantities which are not well known, such as SFR and infall/outflow rates, and often do not address the dynamics within a galaxy. Usually, a galaxy is split into zones as an approximation (e.g., \citealp{timmes1995}). Other models, for example, have calculated the evolution of heavy elemental abundances from C to Zn in our solar neighborhood (e.g., \citealp{chiappini1997, alibes2001, fenner2003, kobayashi2006}). 

Understanding the chemical evolution and isotopic histories of dwarf spheroidal galaxies (dSphs) suffers additional challenges. These systems have a low luminosity because they are gas-poor systems \citep{lanfranchi2003, grebel2003} and have an older stellar population, possibly due to being "fossils" from the pre-reionization era \citep{ricotti2005, theler2019}. These characteristics make it difficult to obtain observational data, which in turn makes it difficult to assess GCE model yields for these galaxies. Furthermore, uncertainties in gas infall/outflow parameterizations also more greatly impact results, as these systems are less gravitationally bound \citep{lanfranchi2003, calura2009}. However, past studies have compiled partial observational elemental abundances for various dSphs (e.g., \citealp{kirby2009, kirby2010, kirby2018, tafelmeyer2010, starkenburg2013, jablonka2015, skuladottir2017, duggan2018, hill2019, skuladottir2019}).

Despite these challenges, attempts have been made to model dSphs. One of the first standardized models for an average dSph was devised by \cite{lanfranchi2003}, who developed a chemical evolution model for 8 dSphs in the Local Group. Other substantial models have generally found that dSphs have a low star formation efficiency, high wind efficiency, low gas content, and therefore, a different chemical evolution history than that of the MW. (e.g., \citealp{recchi2001, venn2004, venn2012, fenner2006, matteucci2008, calura2009, cohen2009, cohen2010, tolstoy2009, kirby2009, kirby2011, kirby2020, venn2012, deboer2012a, deboer2012b, romano2013, leaman2013, homma2014, vincenzo2014, simon2015, theler2019, reichert2020}). Future models hope to provide a better understanding of the chemical evolution differences between dSphs and spiral galaxies, such as the time delay of the Type \RNum{1}a onset and contributions from neutron-capture processes. Additional interest in dSph modeling include, for example, their role in larger galaxy formation \citep{lanfranchi2003, kirby2011}. They are also believed to be the smallest dark matter dominated systems in the Universe \citep{matteucci2008}, and understanding dSphs can further contribute to our knowledge of galactic dark matter profiles \citep{mateo1998, lokas2002, lokas2005, fenner2006, gilmore2007, strigari2018}. Due to their lack of dust and gas as well as large separation between celestial objects, dSphs are also key environments for x-ray and gamma ray studies \citep{jeltema2008, abdo2010}. Finally, dSphs are ideal environments to study the effects of Type \RNum{1}a Supernovae (Type \RNum{1}a SNe) due to their dominance at higher metallicities \citep{kirby2019}.

In this work, we adopt a different approach for describing average isotopic abundances than standard GCE modeling. We follow the methodology taken by \cite{WH13} to model the Sculptor dwarf spheroidal galaxy instead of the MW (see \citealp{battaglia2008} for further background information on Sculptor). The methods used are complementary to GCE models, and other such complimentary models have been explored (e.g., \citealp{ting2012}). The objective is to describe an average isotopic history by scaling isotopic abundances according to the astrophysical processes responsible for their production. We later compare our Sculptor model against the model devised by \cite{WH13} for the MW. This is an improvement over previous studies that interpolate between BBN scaled solar abundances \citep{WH13} or use scaled abundances with $\alpha$-enhancement \citep{limongichieffi2018}.
 
We emphasize that our methodology is distinct from a traditional GCE model. This model can be used for future nucleosynthesis studies (for e.g., as inputs for stellar evolution models). It is a scaling model tuned to data, and provides average isotopic yields at any desired metallicity. In this respect, GCE models are less helpful, as they require inputs from stellar models in the first place and cannot effectively capture large scale stellar formation \citep{prantzos2008}.  

Some benefits with using this approach over a traditional GCE model include:
\begin{enumerate}
    \item This model does not try to describe the more complex principles of GCE, such as the slope of the high-mass end of the stellar IMF and total number of Type \RNum{1}a SNe, which inherently leads to large uncertainties \cite{cote2016}. Consequently, this model does not introduce additional parameters to account for the presence of winds, which is a fundamental characteristic for dSphs (e.g., \citealp{lanfranchi2003}). Rather, the impact of winds is indirectly reflected when the model is fit to observed data and the best-fitting parameter values are found.
    \item GCE models tend to also begin with large uncertainties as a consequence of the first generation of stars not being homogeneously mixed \citep{prantzos2008}. Since GCE models try to improve upon earlier timesteps, they perform poorly at BBN compositions and better near the late-stage evolution composition \citep{WH13, welsh2020}. This model begins where abundances are best known and is then tuned to observed data, so it performs well in describing environments outside the solar neighborhood if sufficient data exists.
    \item The model avoids the discrete, iterative time steps that define GCE models, and is 
    therefore a more comprehensive model throughout all metallicities. Furthermore, this approach avoids the dynamic and evolutionary processes employed in traditional GCE models \citep{WH13}. As stated in \cite{WH13}, stellar yields that are calculated by GCE models are not just based on metallicity but also environment, time, and space which is difficult to assess.
    
\end{enumerate}

Some limitations with using this approach over a traditional GCE model include:
\begin{enumerate}
    \item Our model used the late-stage evolution (LSE) isotopic abundance pattern of the studied system for a fixed point, and a complete set can be difficult to obtain except for compositions similar to the Sun.
    \item The model does not attempt to do GCE calculations to evolve abundances, and instead only uses the core principles of primary and secondary processes as a motivation. Namely,
    \begin{equation}
    \dot X_{\mathrm{\mathit{i,}\mathrm{p}}}=\text{const.}\Rightarrow X_{i,\mathrm{p}}\propto Z,
    \end{equation}
    \begin{equation}
    \dot X_{\mathit{\mathit{i},}\mathrm{s}}\propto X_{\mathit{\mathit{j},}\mathrm{p}}\Rightarrow X_{i,\mathrm{s}}\propto Z^{2},
    \end{equation}
    where $X_{i,\mathrm{p}}$ is the abundance of some isotope $\mathrm{\mathit{i}}$
    made in a primary process, the subscript $\mathrm{s}$ denotes secondary process, and $Z$ is the total metallicity. Since it does not attempt to model dynamic quantities parameterized in GCE models, it avoids the corresponding uncertainties, but at the cost of being able to offer predictions. This more empirically-based method, in addition to isotopic inputs, can still provide consistency checks and suggest trends within the constraints of data to be investigated in future projects.
\end{enumerate}
 
We have chosen the Sculptor dSph for this work due to its significant observational datasets relative to other dSphs, especially in the metal-poor regime \citep{romano2013, jablonka2015} and it has a SFH that appears typical for dSphs \citep{fenner2006, hill2019} so that completed model may characterize the average isotopic history for dSphs. Due to the abundant data, Sculptor is a prime example for dSph research as well as galaxy formation and dark matter studies. Below, we summarize relevant differences in evolution and nucleosynthesis trends between the MW and dSphs.  

Past chemical evolution models suggest dSphs have a high wind efficiency \citep{lanfranchi2003}. This wind can transport enriched gas outside the galaxy, thereby lowering the total gas content and the SFR. Furthermore, the lower gas content suggests a lower metallicity distribution function (MDF) for dSphs which can lead to a broad metal-poor tail (see Figure 3 of \citealp{lanfranchi2003} and Figure 9 of \citealp{homma2014}).

Iron production begins to dominate with the onset of Type \RNum{1}a SNe at an observed metallicity of [Fe/H] $\approx -1$ in the MW \citep{smecker-hane1992, venn2012, kobayashi2011, kobayashi2020}, which is also consistent with results from \cite{WH13}. The Type \RNum{1}a onset occurs at lower metallicities ($-1.5 \leq$ [Fe/H] $\leq -2.0$) in dSphs (e.g., \citealp{recchi2001, lanfranchi2003, fenner2006, tolstoy2009, cohen2009, kirby2011, venn2012, deboer2012a, deboer2012b, romano2013, vincenzo2014, theler2019, reichert2020}). Type \RNum{1}a SNe have also been found to be more dominant in dSphs at higher metallicities than the MW \citep{kirby2020}, which suggests that the decrease of the [$\alpha$/Fe] ratio is steeper than the corresponding decrease in the MW \citep{recchi2001, matteucci2008, vincenzo2014, theler2019}. \cite{tolstoy2009} also found a [Na/Fe] $-$ [Ni/Fe] correlation in dSphs which was first suggested by \cite{nissen1997} who discovered that stars in the MW halo that were deficient in $\alpha$ elements were also deficient in Na, and to a lesser extent, Ni.

It has also been suggested that AGB stars begin dominating the chemical evolution of dSphs at the same time as Type \RNum{1}a SNe \citep{venn2012, theler2019, Skuladottir2020}. Furthermore, the contribution of AGB stars is believed to be stronger in Sculptor than in the MW, especially at higher metallicities \citep{deboer2012a, Skuladottir2020}. The \emph{r}-process to \emph{s}-process transition has been proposed to begin at [Fe/H] = $-$2.6 to [Fe/H] = $-$2 in the MW \citep{truran2002, simmerer2004, kobayashi2011}, but the same transition is still debated for dSphs. Perhaps the most quantitative method to determine this transition was done by \cite{reichert2020} who found that [Eu/Ba] decreased at a metallicity of [Fe/H] = $-$1.57 for Sculptor. The observed [Eu/Fe] data trends in Sculptor have been shown to be similar to the MW, suggesting the \emph{r}-process contribution is similar in these galaxies. In particular, [Eu/Fe] begins to decrease at [Fe/H] $\geq$ $-$1, which is congruent with the Type \RNum{1}a onset value \citep{Skuladottir2020}. Finally, there is evidence that suggests that CCSNe, and not NSM, are the dominant \emph{r}-process progenitors in dSphs \citep{tolstoy2009, duggan2018, skuladottir2019, reichert2020, Skuladottir2020}. We later examine these proposed differences between spiral and dSphs in this work. 

This paper has the following outline: Section \ref{Model Description} highlights differences in the implementation of this model to Sculptor compared to that done in \cite{WH13}. The process for estimating Sculptor's LSE isotopic abundance pattern is also explained. In Section \ref{Fitting Scaling Model to Observational Data}, the model is fit to available data and the best fitting parameters are calculated. Section \ref{Results and Discussion} presents the results obtained from the elemental model and compares it with other dSph studies as well as results from the MW model by \cite{WH13}. The results of the model are also compared with results from a traditional GCE model. Finally, Section \ref{Conclusions} highlights possible conclusions of this work.

\section{Model Description}
\label{Model Description}
We present a model that describes an average isotopic history of the Sculptor dwarf spheroidal galaxy for contributions from massive stars, Type \RNum{1}a SNe, and select neutron-capture processes. We use a similar methodology to \cite{WH13} and describe below only the key differences for its adaptation to a dSph. We also explain the process for estimating Sculptor's LSE composition in Section \ref{Late Stage Evolution Composition}. More information on the methodology employed in this work can be found in \cite{WH13}.

\subsection{Model Differences}
\label{Model Differences}
Here we summarize the differences between our method and that used by \cite{WH13}. The isotopic solar abundance pattern used in this work was taken from a new data set by \cite{Lodders2020} which gives updated values relative to their 2010 publication used in \cite{WH13} \citep{lodders2010}. We also used theoretical BBN abundances of D $= 2.59 \cdot 10^{-5}$, $^3$He $= 0.996 \cdot 10^{-5}$, $^7$Li $= 4.648 \cdot 10^{-10}$, and $^6$Li $= 1.288 \cdot 10^{-14}$ from \cite{cyburt2016}, which gives updated values relative to their 2001 publication used in \cite{WH13} \citep{cyburt2001}. We also used updated values for Y$_\text{P}$ = 0.2470 from \cite{pitrou2018}. We assumed that GCR, classical novae, and the \emph{$\nu$}-process behave similarly in both the MW and Sculptor, as these processes have not been well-explored in dSphs.

\subsubsection{Massive Stars \label{Massive Stars}}

We used yields from \cite{heger2010} for an intermediary point at [Fe/H] = $-$3 found with a Salpeter IMF, a low mixing parameter of 0.02512 used in a running boxcar method, an explosion energy of E = 10$^{51}$ erg, and stellar mass bounds of $10 - 110\,\mathrm{M}_{\odot}$. Since the \cite{heger2010} simulation is typically used for MW studies, we briefly validate our choice for using these massive star yields in this section, and examine the effect of using a non-rotating model at low metallicities later in Section \ref{Impact of Rotation on the Model}. \cite{kirby2019} compared massive star theoretical yields of 3 $\alpha$-elements (Mg, Si, and Ca) and 3 iron-peak elements (Cr, Co, and Ni) from simulations by \cite{nomoto2006} and \cite{heger2010} to their inferred yields in five different dSphs. They determined that the inferred yields of the $\alpha$ and iron-peak elements generally agreed with both simulations for numerous dSphs. Their findings on the Sculptor dSph are summarized below (for more information, see Figure 7 of \citealp{kirby2019}):

\begin{enumerate}
  \item The [Mg/Fe] value agrees with \cite{nomoto2006}, but this is expected as the [Mg/Fe] ratio was chosen as the prior and the [Mg/Fe] yields from \cite{nomoto2006} normalized to quantify the fraction of elements produced in Type \RNum{1}a SNe
  \item The [Si/Fe] value generally agrees with both simulations
  \item The [Ca/Fe] value falls closer to \cite{nomoto2006}
  \item The [Cr/Fe] ratio was in poor agreement with both simulations
  \item The [Co/Fe] value exceeds both simulations but is closer to \cite{heger2010}
  \item The [Ni/Fe] ratio is closer to \cite{heger2010}
\end{enumerate}

It is interesting to note that for the Sculptor dSph, the simulation by \cite{nomoto2006} seems to present more accurate results for the $\alpha$-elements whereas the simulation by \cite{heger2010} performs better for the iron-peak elements. Both perform equally well for different elemental ratios; however, we decided to use the simulation by \cite{heger2010} since we normalize the yields to the observed Fe abundance at [Fe/H] = $-$3.

We used the Salpeter IMF (with $\gamma$ = 1.35) as an input for the massive star fixed point, which has found reasonable success in past chemical evolution models for dSphs \citep{salpeter1955}. Most notably, \cite{lanfranchi2003} explored three distinct IMFs: the Salpeter, a flat Salpeter ($\gamma$ = $-$1.1), and Scalo and found that Salpeter fit the observed properties of dSphs whereas the flat Salpeter and Scalo overproduced and underproduced the alpha ratios, respectively. \cite{romano2013} used the Salpeter IMF and Chabrier-like IMF with $\gamma$ = 1.7 for m $>$ M$_\odot$ from \cite{chabrier2003}. The Salpeter produced reasonable results and modifying the model to the Chabrier IMF only produced miniscule secondary effects. \cite{vincenzo2014} adopted the Salpeter IMF for all their models and also tested an IMF with a steeper slope but found negligible differences. The Salpeter IMF has also been successful in describing observations in other studies \citep{wyse1999, recchi2001, kirby2019}.

It must be noted, however, that there is some degree of uncertainty with using specific IMFs for galaxies other than the MW \citep{matteucci2008}. Indeed, some studies have not had success with the Salpeter IMF. In particular, \cite{calura2009} found better agreement with an IMF slope of $\gamma$ = 1 for numerous dSphs. Additionally, the three component IMF from \cite{kroupa1993} has also been used successfully by other groups \citep{fenner2006, kirby2011, homma2014}.

In this work, we chose the Salpeter IMF due to its reasonable success with past chemical evolution models, though we recognize that no firm consensus exists for the best IMF in dSphs. We accept that there may be other IMFs that provide better results, but currently see no firm consensus about the preferred IMF of choice for dSphs.

\subsubsection{Type \RNum{1}a SNe \label{Type 1a}}

The theoretical Type \RNum{1}a SNe yields and the dominant progenitor must accurately model dSphs. \cite{kirby2019} suggested that sub-Chandrasekhar mass white dwarfs are the dominant Type \RNum{1}a SNe progenitor in dSphs given the comparison to [Ni/Fe] observations (only sub-Chandrasekhar predicted observed values of [Ni/Fe] $<$ $-$0.01).  
\cite{kirby2019} compared three sub-Chandrasekhar models: \cite{leung2020}, \cite{shen2018}, and \cite{bravo2019}. [Ni/Fe] yields especially matched the sub-Chandrasekhar models of \cite{shen2018} and \cite{leung2020} with masses of 1$-$1.10 M$_\odot$, as they were most similar to the observed value of [Ni/Fe] $\approx$ $-$0.3 at [Fe/H] = $-$1.5 for Sculptor. We chose \cite{leung2020} Type \RNum{1}a SNe yields for our model, as Figure 6 of \cite{kirby2019} shows that the standard model of \cite{leung2020} fits Sculptor reasonably well for all elemental ratios. We used their benchmark model 110$-$100$-$2$-$50 (X), corresponding to M$_{\text{WD}}$ = 1.1 M$_\odot$, M$_{\text{He}}$ = 0.1 M$_\odot$, and Z = 0.02. The existence of the sub-Chandrasekhar progenitor was found by \cite{mcwilliam2018} in the Ursa Minor dwarf galaxy (COS 171), and supported by \cite{theler2019} who showed that [Ni/Fe] declined past [Fe/H] $\geq$ $-$2 to subsolar values, implying that the yield of Ni is lower than Fe in Type \RNum{1}a SNe.

Although the choice of the Type \RNum{1}a progenitor is consistent with Sculptor’s observational data, it is important to note that Type \RNum{1}a progenitors likely vary based on metallicity and SFH. Whereas sub-Chandrasekhar mass white dwarfs may be the dominant progenitors of Type \RNum{1}a SNe at low metallicities, near-Chandrasekhar mass white dwarfs may dominate at higher metallicities. Furthermore, [Ni/Fe] evolves differently in galaxies with an extended star formation, such as MW, Fornax, and Leo 1. In such galaxies, it is predicted that they could host Type \RNum{1}a SNe that are delayed \citep{kirby2019}. If single-degenerate near-Chandrasekhar mass white dwarfs are delayed compared to sub-Chandrasekhar mass white dwarfs, it is possible that galaxies with a greater SFH have single-degenerate near-Chandrasekhar mass white dwarfs as their dominant Type \RNum{1}a progenitor, despite trends in [Ni/Fe] (see \citealp{kirby2019} or \citealp{toonen2012} for further discussion).

\subsubsection{Trans-Iron Isotopes \label{Trans-Iron Isotopes}}

Elements that are heavier than iron are not formed by charged particle
fusion reactions because iron has the maximum value for the binding
energy per nucleon \citep{B2FH, bertulani2016}. Ideally, all three components of the \emph{s}-process (ls, hs, and "strong") would be treated separately in the model. Due to the absence of \emph{s}-process contributions for non \emph{s}-only isotopes in our decomposition of Sculptor's LSE and due to lack of observational data, we explicitly model the main \emph{s}-process only. All three components of the main \emph{s}-process (ls, hs, and "strong" components) are treated together and described as a power law:
\begin{equation}
X^{\text{s}}_i(\xi) = X^{\text{s}}_{i, \text{LSE}} \cdot \xi^h \label{s}
\end{equation}
where $X^{\text{s}}_i(\xi)$ is the \emph{s}-process abundance of isotope $i$ as a function of the model parameter, $X^{\text{s}}_{i, \text{LSE}}$ is the \emph{s}-process abundance of isotope $i$ at Sculptor's LSE, and $h$ is a free parameter that describes $\xi$-dependence for a secondary process later fit to observed data (see Section \ref{Fitting Scaling Model to Observational Data} and Table \ref{tab:Parameters}). Since the approximation in Equation \ref{s} follows from lack of observational data to constrain each component separately, results from fitting for the ls and "strong" components are tentative and should be interpreted with care. 

The weak \emph{s}-process should also be modelled separately from the main components. Unfortunately, there is not a clear way to separate these different contributions from the LSE composition given by OMEGA, which mixes them together (see Section \ref{Late Stage Evolution Composition}). One option may be to use the solar abundance composition to approximate the contributions of these processes for all elements (e.g., assume $\approx  10\%$ of Sr is made from the weak \emph{s}-process in the MW, \citealp{WH13}). This method would poorly constrain a proposed weak \emph{s}-process parameter since Sr is dominated by main $s$-process contributions. Moreover, the parameter would only be fit to about $10$ Sr data points. Furthermore, the elemental Sr data average has a high scatter (-2.2 $\leq$ [Sr/Fe] $\leq$ 0.5). We conclude that any resulting fit would likely have negligible benefit given these compounded conditions and, therefore, we choose not to model the weak \emph{s}-process in this work.

We split the \emph{r}-process function into two components to model contributions from CCSNe and NSM. The CCSNe component is described as a power law and the NSM component is described similar to the Type \RNum{1}a SNe function. This choice for NSM follows from an expected delay to contributions consistent with NS merger events in binary systems \citep{mennekens2016}. We define $g$ as the fraction that NSM contribute to the LSE abundance of the \emph{r}-process isotopes: 
\begin{equation}
X^{\text{r}}_i(\xi, g) = g \cdot X^{\text{NSM}}_i(\xi) + (1 - g) \cdot X^{\text{CCSNe}}_i(\xi) \label{r}
\end{equation}
\begin{equation}
X^{\text{CCSNe}}_i(\xi) = X^{\text{r}}_{i, \text{LSE}} \cdot \xi^p \label{CCSNE}
\end{equation}
\begin{equation}
X^{\text{NSM}}_i(\xi) = X^{\text{r}}_{i, \text{LSE}} \cdot \xi \cdot \frac{[\text{tanh}(c \cdot \xi - d) + \text{tanh}(d)]}{[\text{tanh}(c - d) + \text{tanh}(d)]} \label{NSM}
\end{equation}
where Equations \ref{r}, \ref{CCSNE}, and \ref{NSM} follow the same convention as Equation \ref{s}. The parameters $h$, $g$, $p$, $c$, and $d$ are later fit to data. Finally, the $\gamma$-process was chosen to have an equal contribution from primary and secondary seed nuclei:
\begin{equation}
X^{\text{$\gamma$}}_i(\xi) = X^{\text{$\gamma$}}_{i, \text{LSE}} \cdot \xi^{\frac{{h+p}}{2} + 1}
\end{equation}
(see \citealp{WH13} for further discussion).

\subsection{Late Stage Evolution Composition} \label{Late Stage Evolution Composition}

Ideally, the LSE composition would come from observations. Since a complete isotopic abundance set is not observable for Sculptor, we use the OMEGA GCE code \citep{cote2017} to estimate the LSE composition. We used an initial galaxy mass of 7.8x10$^6$ M$_\odot$, a ratio between inflow and outflow rate of 1.02, a ratio between outflow rate and SFR of 8, the number of NSM per stellar mass of 2.0x10$^{-5}$, a mass ejected from NSM of 1.2x10$^{-1}$ M$_\odot$, and computed the GCE model across 10,000 timesteps. The results delineated yields by process for further analysis. We tested several AGB and massive star theoretical yields (with and without rotation) to determine which final computed compositions provided the best fit to Sculptor's LSE, as well as an intermediary point at [Fe/H] = $-$3 (See Section \ref{Impact of Rotation on the Model}). The best fit isotopic abundances were found using inputs from \cite{limongichieffi2018} (see Table \ref{tab:AGB/Massive Star Theoretical Yields}) and a rotational velocity of 300 km s$^{-1}$, which underproduced and overproduced the Ba LSE composition, respectively. Thus, we computed a weighted average between the yields to fix the Ba LSE composition. The AGB and massive star theoretical yields also include contributions from winds. Additional inputs include the Sculptor SFH from \cite{deboer2012a}, Type \RNum{1}a SNe yields from \cite{leung2020} (see Section \ref{Type 1a}), and a Salpeter IMF (See Section \ref{Massive Stars}). For the LSE yields, we extracted isotopic abundances from the GCE model at the timestep corresponding to [Fe/H] = $-$1.25, which denotes the highest metallicity where Sculptor data is plentiful.

\begin{figure*}
\includegraphics[width=6.5in]{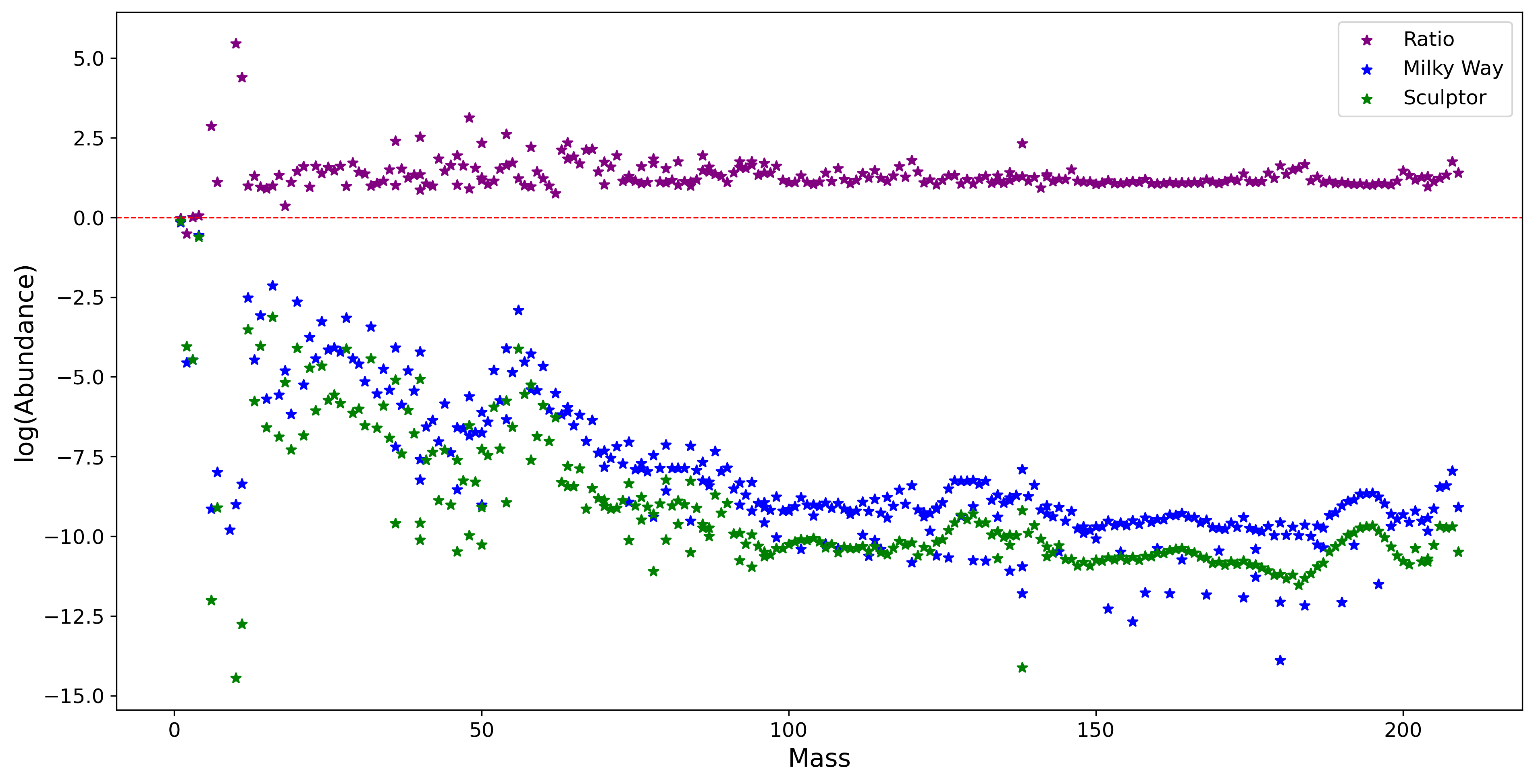}
\caption{Isotopic abundances for the Sun \citep{Lodders2020} and the LSE of Sculptor computed using OMEGA \citep{cote2017}. The ratios between MW and Sculptor abundances are also shown.} \label{fig:Abundances of Late Stage Evolution}
\centering
\end{figure*}

The LSE composition of the MW and Sculptor and the ratio between them are given in Figure \ref{fig:Abundances of Late Stage Evolution}. Both show known characteristic features of nucleosynthesis, with Sculptor abundances consistently below MW abundances by $\approx$ 1.3 dex on average. The abundances deviate most before the iron-peak, which is shown by the larger ratios between A $\sim$ 40 $-$ 50. While there is less deviation between the abundances after the iron-peak, there are some notable exceptions, such as $^{138}$La which is 2.3 dex above the MW composition. The weak \emph{s}-process plateau is comparatively less defined for Sculptor yields. Note that the Sculptor yields omit $p$-isotope contributions from photo-disintegration events, which were not included in the OMEGA model.

\subsection{Impact of Rotation Using OMEGA \label{Impact of Rotation on the Model}}
\begin{figure*}
\includegraphics[width=6.5in]{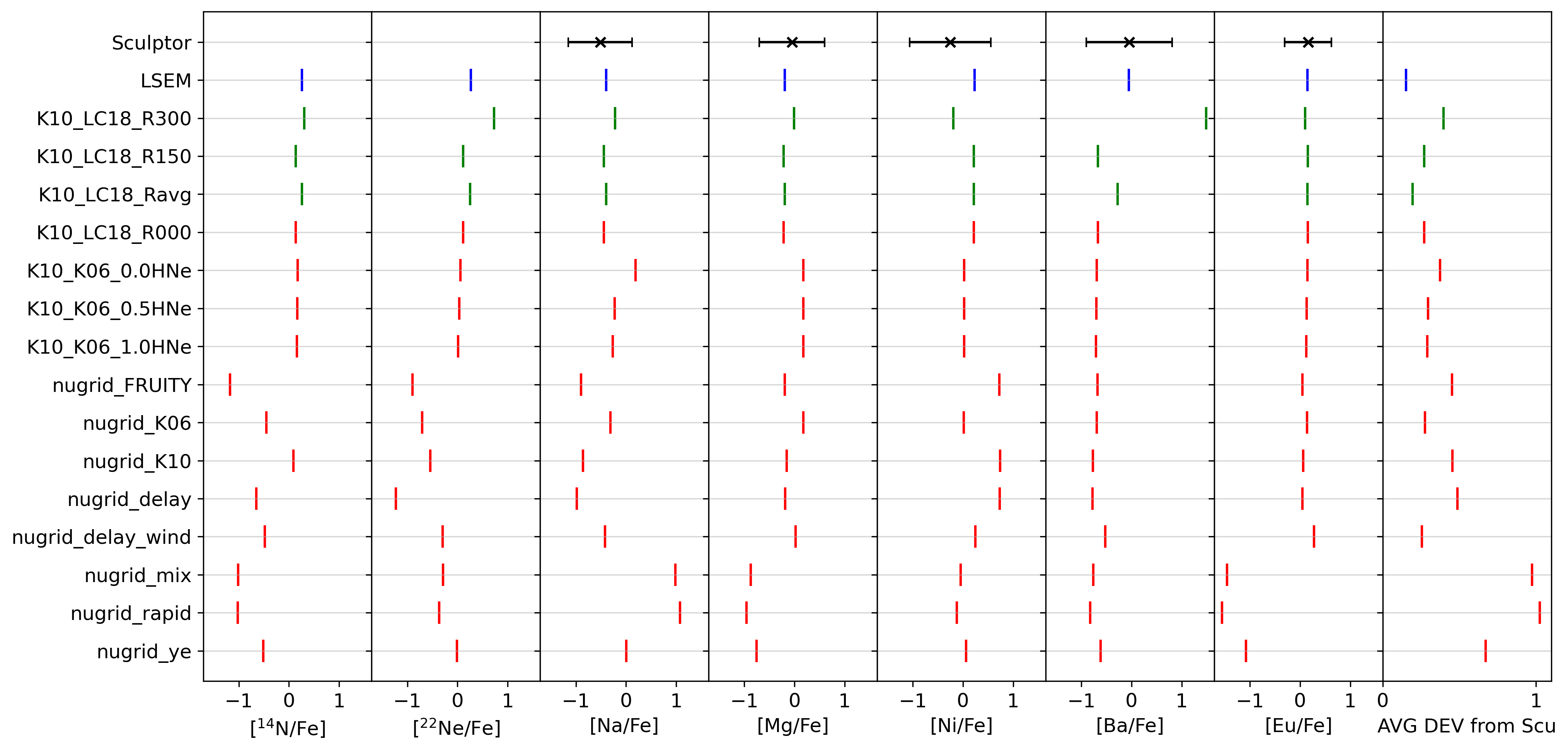}
\caption{A comparison of OMEGA GCE models using rotating and non-rotating theoretical models (described in Table \ref{tab:AGB/Massive Star Theoretical Yields}) with the Sculptor observed average at [Fe/H] = $-$1.25. The rightmost column depicts the average deviation of each model from Sculptor observed averages for [Na/Fe], [Mg/Fe], [Ni/Fe], [Ba/Fe], and [Eu/Fe]. Green and red markers indicate models that do and do not consider rotation, respectively.} \label{fig:Rotation_minus1.25}
\centering
\end{figure*}

\begin{figure*}
\includegraphics[width=6.5in]{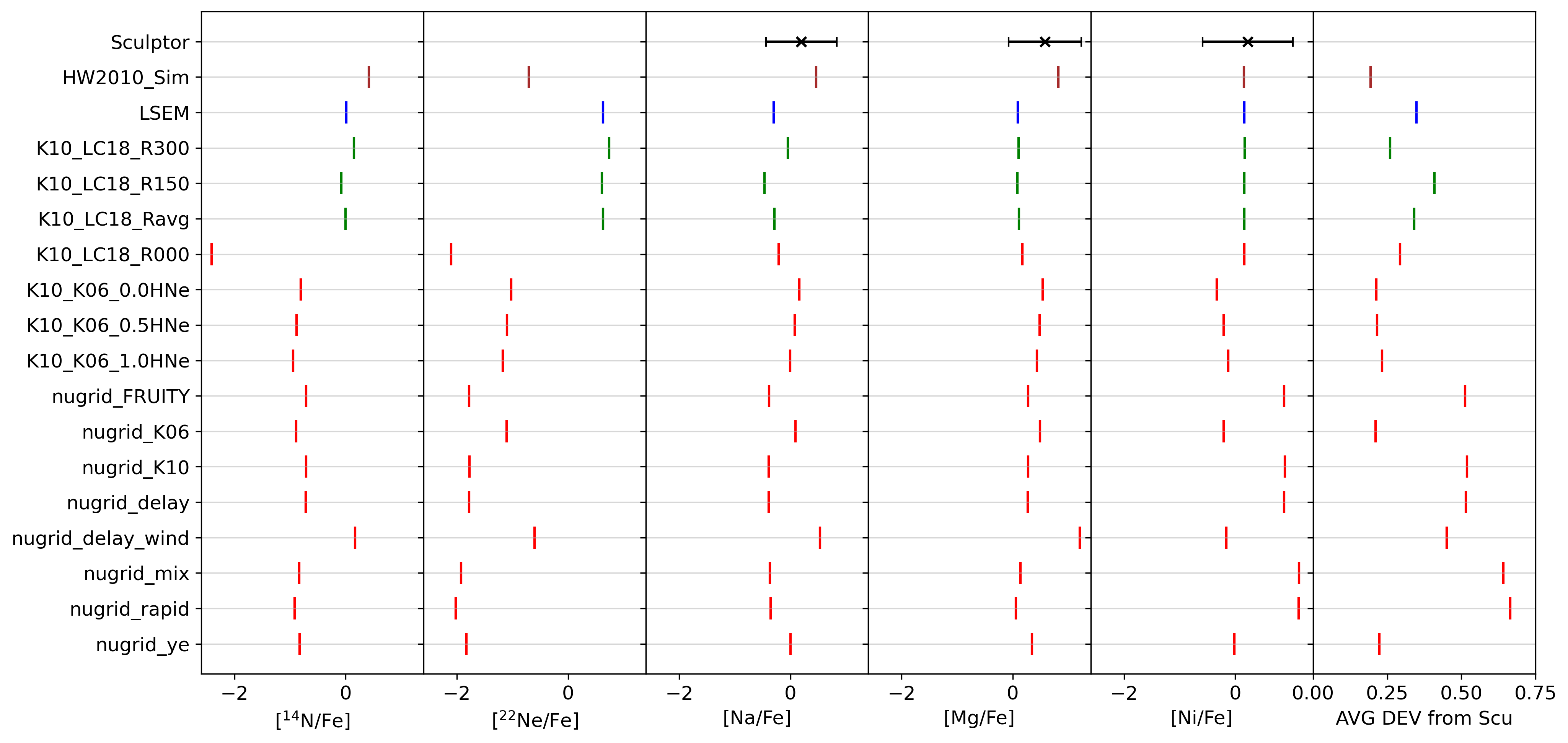}
\caption{A comparison of OMEGA GCE models using rotating and non-rotating theoretical models with the Sculptor observed average at [Fe/H] = $-$3. The figure follows the same convention as Figure \ref{fig:Rotation_minus1.25}. \label{fig:Rotation_minus3}}
\centering
\end{figure*}

The LSE composition used rotating AGB and massive star yields from \cite{limongichieffi2018}, but the intermediate point at [Fe/H] = $-$3 was found with non-rotating massive star yields from \cite{heger2010}. In this section, we validate our choice for using these yields with a different rotation prescription by comparing OMEGA's output from several rotating and non-rotating theoretical yields (shown in Table \ref{tab:AGB/Massive Star Theoretical Yields}) with Sculptor's observed data averages. Each theoretical model was used as input in OMEGA, and the abundances of $^{14}$N, $^{22}$Ne, Na, Mg, and Ni were extracted from OMEGA at [Fe/H] = $-$1.25 and [Fe/H] = $-$3. In addition, [Ba/Fe] and [Eu/Fe] abundances were also extracted at [Fe/H] = $-$1.25. These theoretical values were compared with the Sculptor observed average at the same metallicity to compare theoretical models. The rightmost columns of Figures \ref{fig:Rotation_minus1.25} and \ref{fig:Rotation_minus3} show the average deviation between all theoretical values and the Sculptor observed average for each model considered.

Figure \ref{fig:Rotation_minus1.25} shows that the rotating models and a majority of the non-rotating models are able to reproduce observed Sculptor trends for [Na/Fe] and [Mg/Fe], although most models overproduce [Ni/Fe]. Both rotating and non-rotating models are generally able to reproduce observed [Eu/Fe]. However, almost every theoretical model underproduces [Ba/Fe], with only K10$\_$LC18$\_$R300 overproducing this value by $\approx$ 1.5 dex. We took the closest model to the observed [Ba/Fe], K10$\_$LC18$\_$Ravg, and performed a weighted average between it and K10$\_$LC18$\_$R300. The resulting AGB/massive star yields were used as an input to determine the OMEGA LSE Model (LSEM). The yields from LSEM at [Fe/H] = $-$1.25 determine the LSE isotopic composition.

The LSEM provides the smallest average deviation from Sculptor (Figure \ref{fig:Rotation_minus1.25}). It should be noted, however, that LSEM does not reproduce [Ni/Fe] trends at [Fe/H] = $-$1.25, although the other models considered do not either. It is interesting to see that K10$\_$LC18$\_$R300 is the only model to accurately reproduce the observed [Ni/Fe] Sculptor value, which suggests that rotation may have been an integral part of Sculptor's LSE. This contradicts previous studies, however, that show that in the MW, rotating yields best fit the observed data at the low metallicity region whereas non-rotating yields best fit the data at the high metallicity region \citep{chiappini2006, rizzuti2019}.

Figure \ref{fig:Rotation_minus1.25} also shows that $^{14}$N and $^{22}$Ne are overproduced in rotating models, as suggested from previous studies \citep{prantzos2018}. However, since there is no Sculptor data to compare the results from rotating and non-rotating OMEGA simulations, it is difficult to discern whether a rotating or non-rotating yield input is better for these ratios  in a model that relies upon fits to data.

Whereas Figure \ref{fig:Rotation_minus1.25} shows that LSEM is more accurate than other theoretical models at [Fe/H] = $-$1.25, this does not guarantee that the same model is accurate at the intermediary fixed point ([Fe/H] = $-$3). Figure \ref{fig:Rotation_minus3} compares the same rotating and non-rotating theoretical models to the Sculptor data averages of [Na/Fe], [Mg/Fe], and [Ni/Fe] at [Fe/H] = $-$3, along with the addition of a theoretical non-rotating massive star simulation by \cite{heger2010}. The right-most column shows that HW2010$\_$Sim is the closest model to the Sculptor data averages and performs significantly better than the LSEM. Thus, we perform a massive star and Type \RNum{1}a scaling at the intermediary point of [Fe/H] = $-$3 with this model (see Section \ref{Massive Type1a Scalings}). Once again, these results conflict with \cite{chiappini2006, rizzuti2019}, as they suggest that slow rotating yields best fit MW data at low metallicities, though it is possible that rotation is different between the MW and dSphs. For example, \cite{tsujimoto2006} predicted that massive stars in dSphs have a smaller rotation than in the MW.

\begin{table*}
    \centering
    \caption{\label{tab:AGB/Massive Star Theoretical Yields}Theoretical AGB and massive star theoretical models. All models can be found in the online NuPyCEE package \citep{cote2016}, with the exception of LSEM.}
    \begin{tabular}{*3l}
        \toprule
        Model Acronym & Massive Star Yields & AGB Star Yields \\
        \midrule
        LSEM & Weighted Averages from models K10$\_$LC18$\_$R300 and K10$\_$LC18$\_$Ravg & \cite{karakas2010}\\
        HW2010$\_$Sim & \cite{heger2010} & N/A\\
        K10$\_$LC18$\_$R300 & \cite{limongichieffi2018}, with a rotational velocity of 300 km s$^{-1}$ & \cite{karakas2010}\\
        K10$\_$LC18$\_$R150 & \cite{limongichieffi2018}, with a rotational velocity of 150 km s$^{-1}$ & \cite{karakas2010}\\
        K10$\_$LC18$\_$Ravg & \cite{limongichieffi2018}, with an averaged velocity by \cite{prantzos2018} & \cite{karakas2010}\\
        K10$\_$LC18$\_$R000 & \cite{limongichieffi2018}, with a rotational velocity of 0 km s$^{-1}$ & \cite{karakas2010}\\
        K10$\_$K06$\_$0.0HNe & \cite{kobayashi2006}, with a 0.0 HNe fraction & \cite{karakas2010}\\
        K10$\_$K06$\_$0.5HNe & \cite{kobayashi2006}, with a 0.5 HNe fraction & \cite{karakas2010}\\
        K10$\_$K06$\_$1.0HNe & \cite{kobayashi2006}, with a 1.0 HNe fraction & \cite{karakas2010}\\
        nugrid\_FRUITY & \cite{ritter2018} & \cite{cristallo2009}\\
        nugrid\_K06 & \cite{kobayashi2006} & \cite{ritter2018}\\
        nugrid\_K10 & \cite{ritter2018} & \cite{karakas2010}\\
        nugrid\_delay & \cite{ritter2018}, with delay mass prescription from \cite{fryer2012} & \cite{ritter2018}\\
        nugrid\_delay\_wind & \cite{ritter2018}, with delay mass prescription and pre-explosion wind from \cite{fryer2012} & \cite{ritter2018}\\
        nugrid\_mix & \cite{ritter2018}, with mix of delay and rapid mass prescriptions from \cite{fryer2012} & \cite{ritter2018}\\
        nugrid\_rapid & \cite{ritter2018}, with rapid mass prescription from \cite{fryer2012} & \cite{ritter2018}\\
        nugrid\_ye & \cite{ritter2018}, with mass-cut prescription based on the electronic fraction from \cite{cote2016} & \cite{ritter2018}\\
        \bottomrule
    \end{tabular}
\end{table*}

\subsection{Massive and Type \RNum{1}a SNe Scalings \label{Massive Type1a Scalings}}

We assume that both Type \RNum{1}a SNe and massive stars (which include AGB yields below the iron-peak) contribute a fraction of each intermediate-mass and iron-peak isotope abundance at [Fe/H] = $-$3. We use \cite{heger2010} for massive star yields and \cite{leung2020} for the Type \RNum{1}a SNe yields (see Sections \ref{Massive Stars} and \ref{Type 1a} for specific reasoning behind this choice). We define a fraction, $f$, to be the contribution of Type \RNum{1}a SNe to the observed $^{56}$Fe LSE abundance. Thus, we scale each LSE isotope in $12<A<70$ by the factor: $f$ $\cdot$ X$^{\text{LSE}}_{56}$ $/$ X$^{\text{Ia}}_{56}$, where X$^{\text{LSE}}_{56}$ is the observed LSE abundance of $^{56}$Fe in Sculptor and X$^{\text{Ia}}_{56}$ is the Type \RNum{1}a contribution taken from the OMEGA yields. This shifts the isotopic abundance pattern for Type \RNum{1}a SNe at [Fe/H] = $-$3 to fit the yield for $^{56}$Fe. The fraction $f$ is a free variable that is later fit to observational data (see Table \ref{tab:Parameters}). The OMEGA LSE massive star yields were scaled in a similar fashion by a factor (1 $-$ $f$) $\cdot$ X$^{\text{LSE}}_{56}$ $/$ X$^{\text{massive}}_{56}$, where X$^{\text{LSE}}_{56}$ is the OMEGA LSE abundance of $^{56}$Fe in Sculptor and X$^{\text{massive}}_{56}$ is the massive contribution. The effect of this scaling is to normalize the Fe yields for both massive stars and Type \RNum{1}a SNe to its observed value.

\begin{figure*}
\includegraphics[width=6.5in]{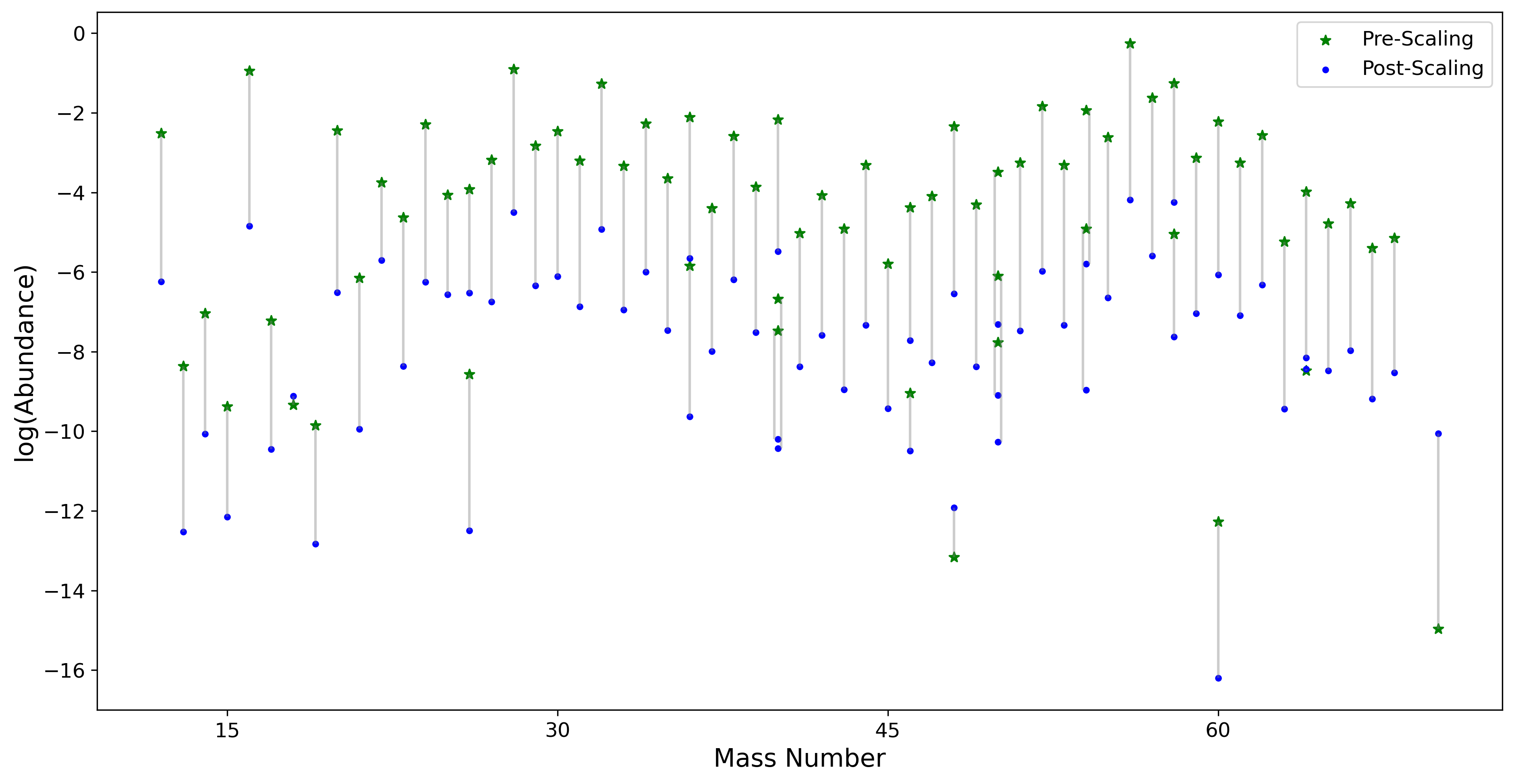}
\caption{The Type \RNum{1}a yields from \citealp{leung2020} scaled to a nominal value of $^{56}$Fe$_{\text{\RNum{1}a}}$/$^{56}$Fe$_{\text{LSE}}$ = 0.85. Shown are the yields before (Stars) and after (Circles) fitting against massive star yields.} \label{fig:type1aScalings}
\centering
\end{figure*}

\begin{figure*}
\includegraphics[width=6.5in]{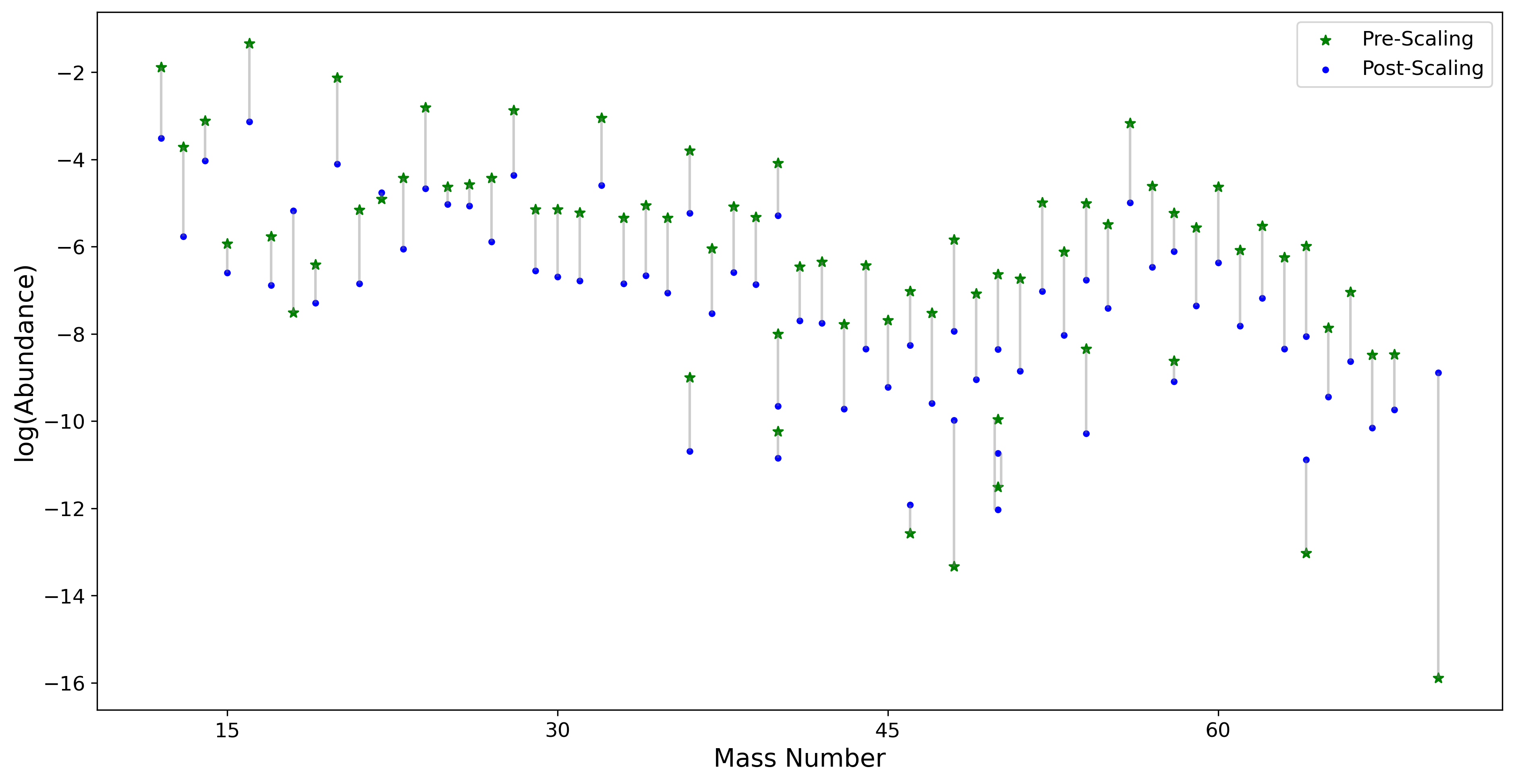}
\caption{The Massive star yields from \citealp{heger2010} fitted to observed Sculptor data and scaled to a nominal value of $^{56}$Fe$_{\text{massive}}$/$^{56}$Fe$_{\text{LSE}}$ = 0.15. Shown are the yields before (Stars) and after (Circles) fitting against Type \RNum{1}a yields. \label{fig:massiveScalings}}
\centering
\end{figure*}

An additional scaling was performed to ensure that all other isotopic contributions from both massive stars and Type \RNum{1}a SNe summed to the LSE abundance for each isotope between $12 \leq A \leq 70$:

\begin{equation}
X^{\text{massive}}_{i, f} = \Bigg( \ddfrac{X^{\text{LSE}}_i}{X^{\text{massive}}_{i, 0} + X^{\text{a}}_{i, 0}}\Bigg)X^{\text{massive}}_{i, 0}
\end{equation}

\begin{equation}
X^{\text{\RNum{1}a}}_{i, f} = \Bigg( \ddfrac{X^{\text{LSE}}_i}{X^{\text{massive}}_{i, 0} + X^{\text{\RNum{1}a}}_{i, 0}}\Bigg)X^{\text{\RNum{1}a}}_{i, 0} 
\end{equation}
where X$_{i, 0}$ and X$_{i, f}$ are the initial and scaled abundances of isotope \textit{i}, respectively, and $X^{\text{LSE}}_i$ is the observed LSE abundance of isotope \textit{i} given by OMEGA. Whereas the abundance pattern within each model is distorted due to the scaling processes, the isotopic ratios are preserved across each model. For example, the even-odd effect is still apparent, and the iron-peaks for both massive (Figure \ref{fig:massiveScalings}) and Type \RNum{1}a (Figure \ref{fig:type1aScalings}) can be seen.

\section{Fitting Scaling Model to Observational Data}
\label{Fitting Scaling Model to Observational Data}
The isotopic scaling functions were summed into elemental scaling functions and fit to Sculptor stellar abundance data to tune the free parameters. We used data from \cite{kirby2009} for $\alpha$ elements (Mg, Si, Ca, and Ti) and Fe abundances for almost 400 stars measured using Keck/DEIMOS medium-resolution spectra. We used data from \cite{tafelmeyer2010} for iron-peak, neutron-capture, and $\alpha$ elements for two metal-poor stars using high spectroscopic resolution with VLT/UVES. We used data from \cite{starkenburg2013} for $\alpha$ and iron-peak elements for seven low-metallicity stars using ESO VLT. We used data from \cite{hill2019} for $\alpha$ elements (O, Mg, Si, Ca, Ti), iron-peak elements (Sc, Cr, Fe, Co, Ni, Zn), and neutron-capture elements (Ba, La, Nd, Eu) for 99 red-giant branch stars using high-resolution VLT/FLAMES spectroscopy. We used data from \cite{skuladottir2017} for Zn abundances for approximately 100 red giant branch stars using ESO VLT/FLAMES/GIRAFFE spectra. We used data from \cite{kirby2018} for iron-peak elements (Cr, Co, Ni) for approximately 300 red giant stars using Keck/DEIMOS medium-resolution spectra. We used data from \cite{duggan2018} for [Ba/Fe] for 120 red giant branch stars in Sculptor using medium-resolution spectroscopy. We used data from \cite{skuladottir2019} for neutron-capture elements Y, Ba, La, Nd, and Eu for 98 stars using ESO VLT/FLAMES spectre. We used data from \cite{jablonka2015} for iron-peak, neutron-capture, and $\alpha$ elements for 5 metal-poor stars using the medium-resolution CaT from \cite{battaglia2008}. All data is given in units of [X/Fe]. 

The fitting procedure was implemented as follows: the [Fe/H] axis was split into 100 bins in the range $-$5 to $-$1. Similarly, the [X/Fe] axis was split into 100 bins in the range of the elemental data. Every data point was assigned a Gaussian distribution using their respective observational uncertainties:
\begin{equation}
f_i = \text{exp}\Bigg[ -0.5 \cdot \Bigg(\frac{x_i - x_0}{\sigma_x}\Bigg)^2 + -0.5 \cdot \Bigg(\frac{y_i - y_0}{\sigma_y}\Bigg)^2 \Bigg]
\end{equation}
where $f_i$ is the value of the distribution at bin ($x_i$, $y_i$) on the [Fe/H] and [X/Fe] axes, $x_0$ and $y_0$ are the data point values, and $\sigma_x$ and $\sigma_y$ are the associated uncertainties for each $x_0$ and $y_0$.  The distribution values were then averaged for each bin and a standard deviation was found. A differential evolution algorithm \citep{storn1997} was implemented to minimize the $\chi^2$ between the average observed curve and the functional curve while preserving the expected chemical trends to avoid over-fitting. The free parameter values that resulted in the smallest $\chi^2$ values for Mg, Ba, Eu, and Sr are shown in Table \ref{tab:Parameters}. 

\subsection{Type \RNum{1}a Parameters}
\label{Type1a Parameters}
\begin{figure*}
\includegraphics[width=6.5in]{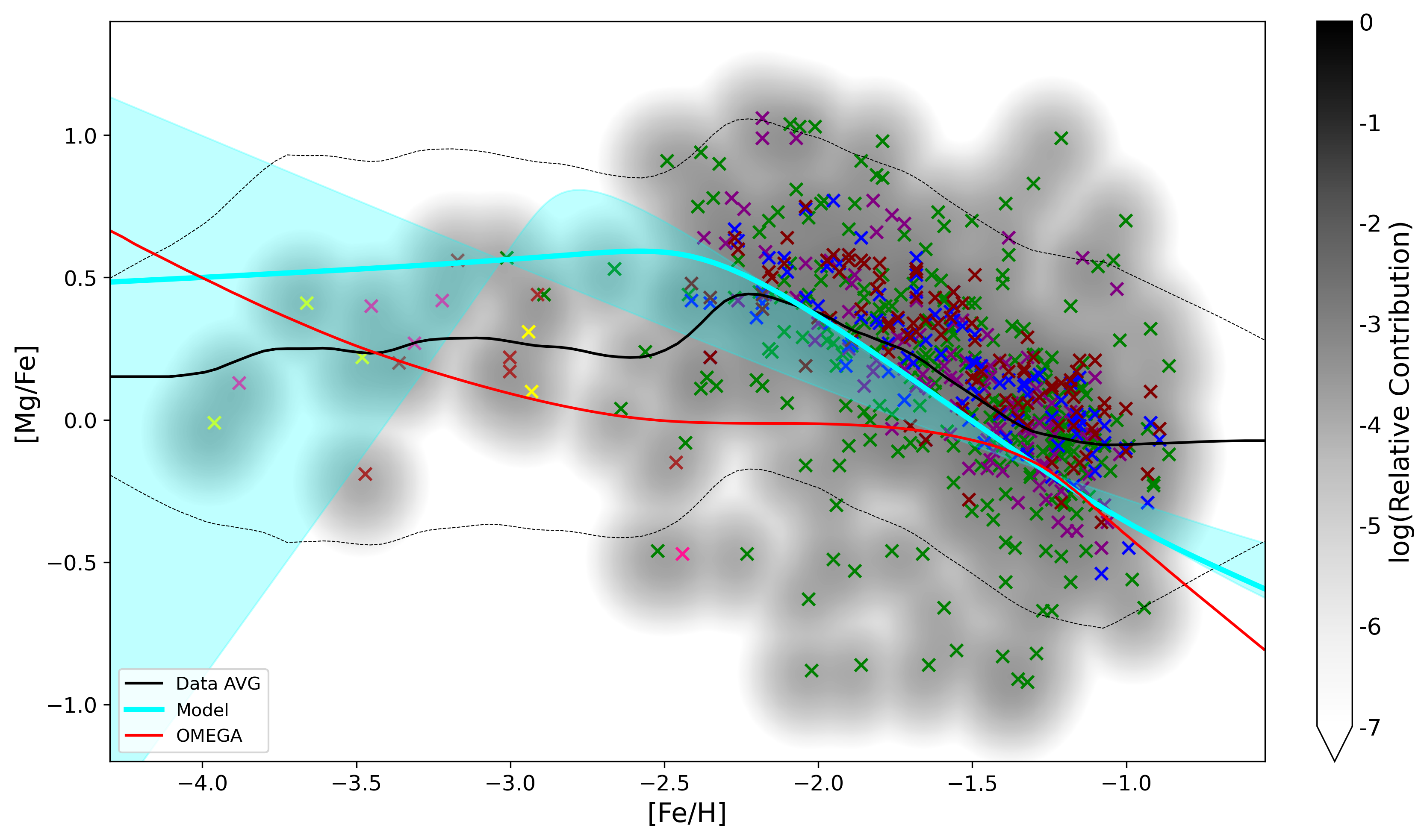}
\caption{The resulting model for [Mg/Fe] determined after finding the best-fitting parameters. The dark shadow background shows the Gaussian distributions of stellar data according to the uncertainties. The solid black line gives the average curve, with the dotted black lines showing the standard deviation curves. The model line is shown by the solid cyan line, and the OMEGA GCE model is shown by the solid red line. The cyan background behind the model line covers the range of possible model curves across the complete parameter space for $a$, $b$, and $f$. Data points are taken from \citealp{kirby2009} (green), \citealp{tafelmeyer2010} (yellow), \citealp{starkenburg2013} (brown), \citealp{jablonka2015} (pink), \citealp{kirby2018} (purple), \citealp{skuladottir2019} (maroon), and \citealp{hill2019} (blue).}
\label{fig:Mg}
\centering
\end{figure*}

The Type \RNum{1}a SNe parameters $a$, $b$, and the fraction of $^{56}$Fe$_{\text{LSE}}$ attributed to Type \RNum{1}a SNe, $f$, were found by using [Mg/Fe] due to the prevalence of data available. Type \RNum{1}a SNe and massive stars both contribute significantly to the evolution of Mg and Fe \citep{heger2010, kirby2019}. The parameters were chosen to minimize the $\chi^2$ between the [Mg/Fe] model and the observed [Mg/Fe] data averages. The best fitting parameters were found to be $a$ = 1.628, $b$ = 1.073, and $f$ = 0.863 with a $\chi^2$ of 0.359. The resulting model for [Mg/Fe] is shown in Figure \ref{fig:Mg}, along with data and the OMEGA evolution for comparison.

The model curve tracks the data average. At metallicities below Type \RNum{1}a onset ([Fe/H] $< \approx$ $-$2.5), magnesium and iron scale together since only massive stars contribute to their production. Here, the model line sits about 0.3 dex above the average line. This discrepancy is a consequence of the fixed point obtained at [Fe/H] = $-$3 using \cite{heger2010} (see Section \ref{Massive Stars}). Whereas the OMEGA GCE evolution fits marginally closer to the data at this metallicity, using it as the fixed point instead would drive the model below the data average at higher metallicities. We choose to prioritize fits to higher metallicities, where data is more plentiful. We also note the paucity of data below [Fe/H] = $-$2.5. It is reasonable to predict that the [Mg/Fe] trend would plateau in this regime, given their production mechanism, though the value of that plateau is poorly constrained. Indeed, the model may not be a good representation of the average chemical evolution as metallicities decrease substantially below [Fe/H] $\approx$ $-$3. The chemical evolution may have been more stochastic in this regime, as some studies have explored for the MW halo \citep{welsh2020}.

The drop of the curve to LSE values at [Fe/H] $\approx$ $-$2.5 is caused by the Type \RNum{1}a onset which coincides with iron production from Type \RNum{1}a SNe and drives the ratio down to its late stage value. The model line mirrors the average line after the drop up to [Fe/H] = $-$1.25, and the slopes of both curves are similar during the onset. 

Although the data average plateaus after LSE values ([Fe/H] $>$ $-$1.25), the model does not. There is no metal-rich data available above [Fe/H] $\approx$ $-$1 to constrain [Mg/Fe], so caution should be used with the model in this regime. If Sculptor continued to have active star formation, then the model might capture an accurate data average trend for some finite range above [Fe/H] = $-$1, as CCSNe continued to expel material that escapes the gravity well, while Type \RNum{1}a SNe injects the ISM with a lower [Mg/Fe] ratio.   

\subsection{\emph{r}-process and \emph{s}-process Parameters}
\label{r-process and hs Paramaters}
\begin{figure*}
\includegraphics[width=6.5in]{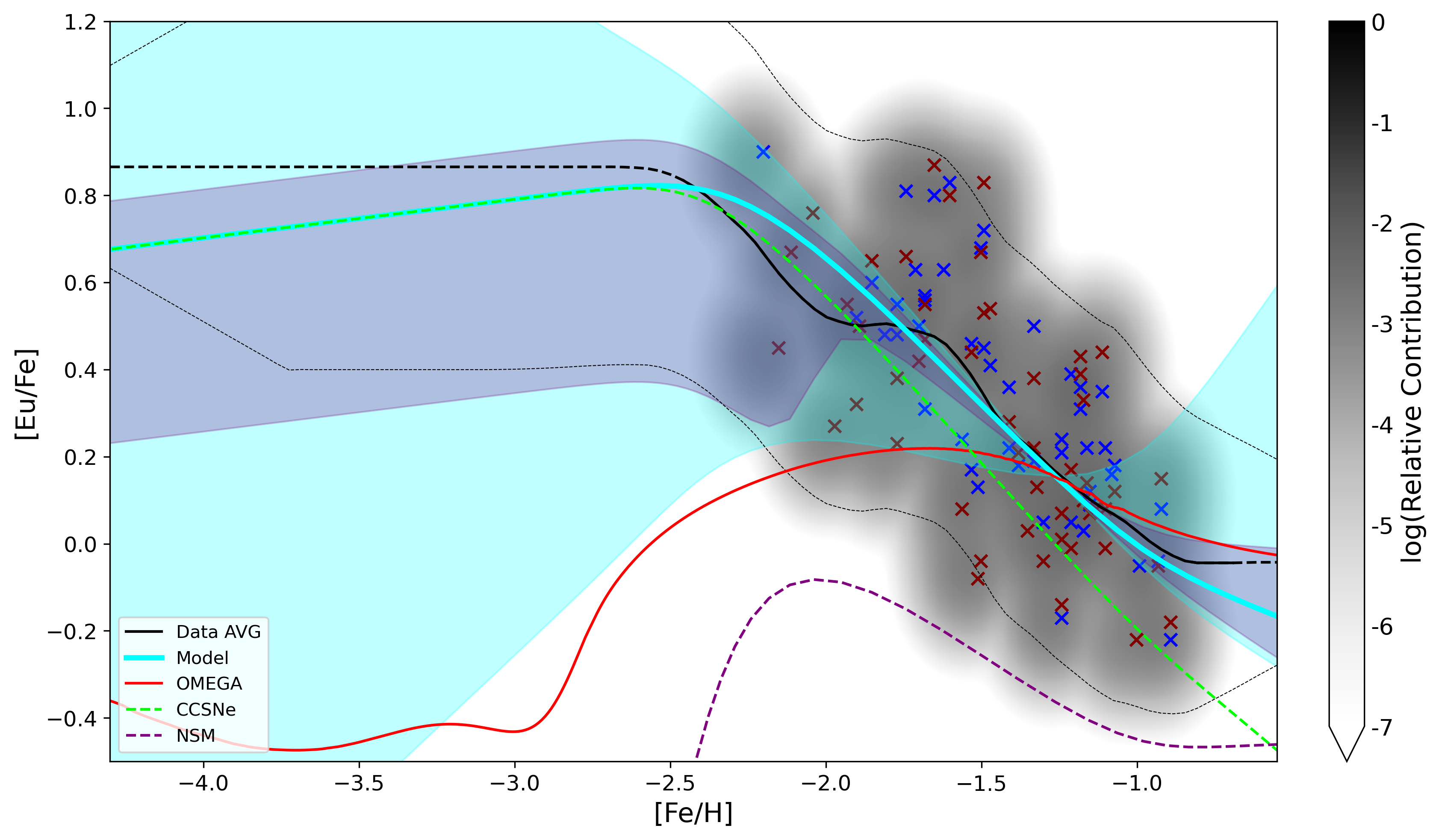}
\caption{The resulting model for [Eu/Fe] determined after finding the best-fitting parameters. The figure follows the same convention as Figure \ref{fig:Mg}, with the addition of the CCSNe (green dashed line) and NSM (purple dashed line) contributions to the model shown separately. The cyan background behind the model line covers the range of possible model curves across the complete parameter space for $c$, $d$, $p$, and $g$, and the darker purple hue shows the subset of this range that is constrained by the parameter $g$. Plotted data is from \citealp{skuladottir2019} (maroon) and \citealp{hill2019} (blue).} \label{fig:Eu}
\centering
\end{figure*}

Four parameters are required to describe the \emph{r}-process scaling: the primary process exponent ($p$), the fraction that NSM contribute to \emph{r}-process isotopes ($g$), as well as the scaling and shifting factors of the tanh function to describe contributions from NSM ($c$ and $d$, respectively). Eu was chosen for fitting these parameters as it is an \emph{r}-process peak element with two isotopes, $^{151}$Eu and $^{153}$Eu, which both have dominant (85 per cent of its solar value, \citealp{WH13}) contributions from the \emph{r}-process. The best fitting parameter values were found to be $p$ = 0.391, $g$ = 0.303, $c$ = 5.343, and $d$ = 0.450 with $a$ $\chi^2$ = 0.022. The previously found best fitting values for $a$, $b$, and $f$ (from Section \ref{Type1a Parameters}) were used for Fe. Whereas a nominal value of $h$ = 1.5 was used for the \emph{s}-process parameter in \cite{WH13}, the LSE of Sculptor computed by OMEGA gives no \emph{s}-process contributions for Eu. The best fitting functional curve and average curve for [Eu/Fe] are shown in Figure \ref{fig:Eu}. 

\begin{figure*}
\includegraphics[width=6.5in]{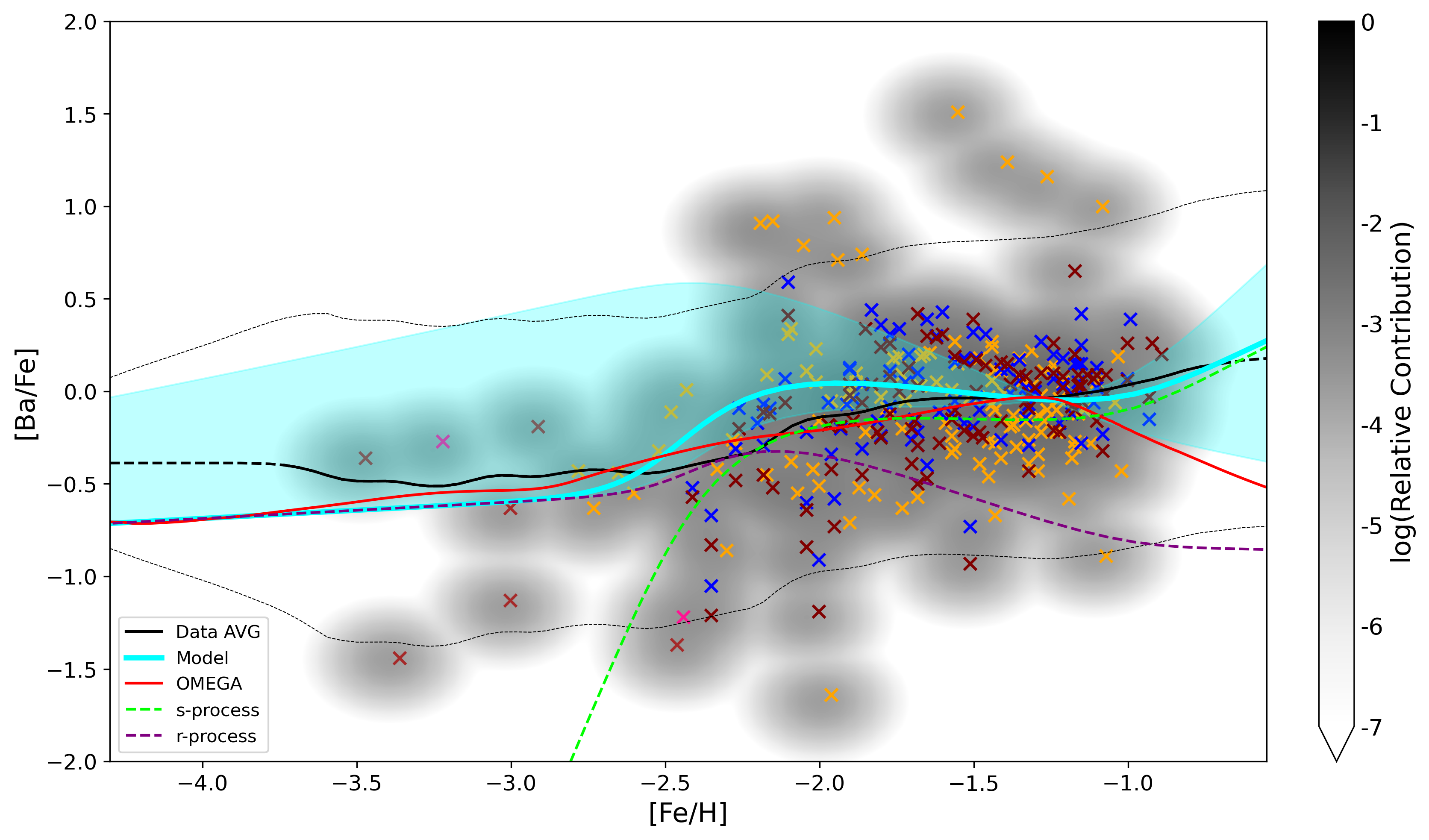}
\caption{The resulting model for [Eu/Fe] determined after finding the best-fitting parameters. The cyan background behind the model line covers the range of possible model curves across the parameter space for $h$. The figure follows the same convention as Figure \ref{fig:Mg}, with the addition of the \emph{s}-process (green dashed line) and \emph{r}-process (purple dashed line) contributions to the model shown separately. Plotted data is from \citealp{starkenburg2013} (brown), \citealp{jablonka2015} (pink), \citealp{duggan2018} (orange), \citealp{skuladottir2019} (maroon), and \citealp{hill2019} (blue).} \label{fig:Ba}
\centering
\end{figure*}

To determine the best value for the \emph{s}-process parameter $h$, Ba was chosen as it has two s-only isotopes, $^{134}$Ba and $^{136}$Ba, along with three isotopes, $^{135}$Ba, $^{137}$Ba, and $^{138}$Ba, which have contributions from both the \emph{s}-process and \emph{r}-process \citep{WH13}. The $\gamma$-process contributions from the $p$ isotopes $^{130}$Ba and $^{132}$Ba were not considered for this fit, and are negligible at all metallicities \citep{WH13}. The best fitting parameter value was found to be $h$ = 1.798 with $a$ $\chi^2$ = 0.448. The previously found best fitting values for $a$, $b$, $f$ were used for Fe, and those for $p$, $g$, $c$, and $d$ were used for the \emph{r}-process contributions to Ba. The best fitting model for [Ba/Fe] is shown in Figure \ref{fig:Ba}.

\section{Results and Discussion}
\label{Results and Discussion}

\begin{figure*}
\includegraphics[width=6.5in]{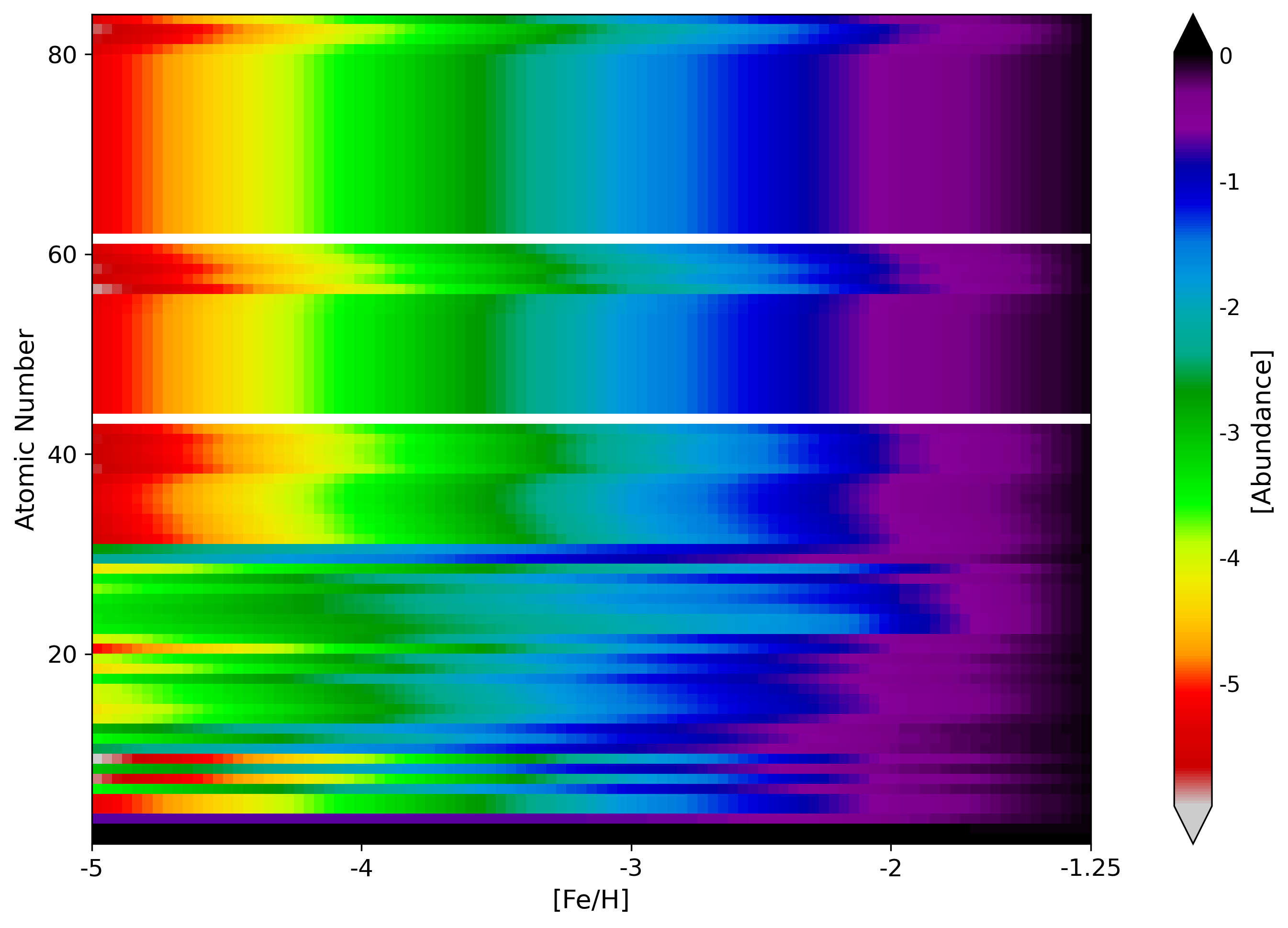}
\caption{The complete elemental scaling of the model. The abundances are given relative to their MW solar values. The model only considers \emph{s}-process contributions to the \emph{s}-process peak elements. \label{fig:Scalings}}
\centering
\end{figure*}

\begin{figure*}
\includegraphics[width=6.5in]{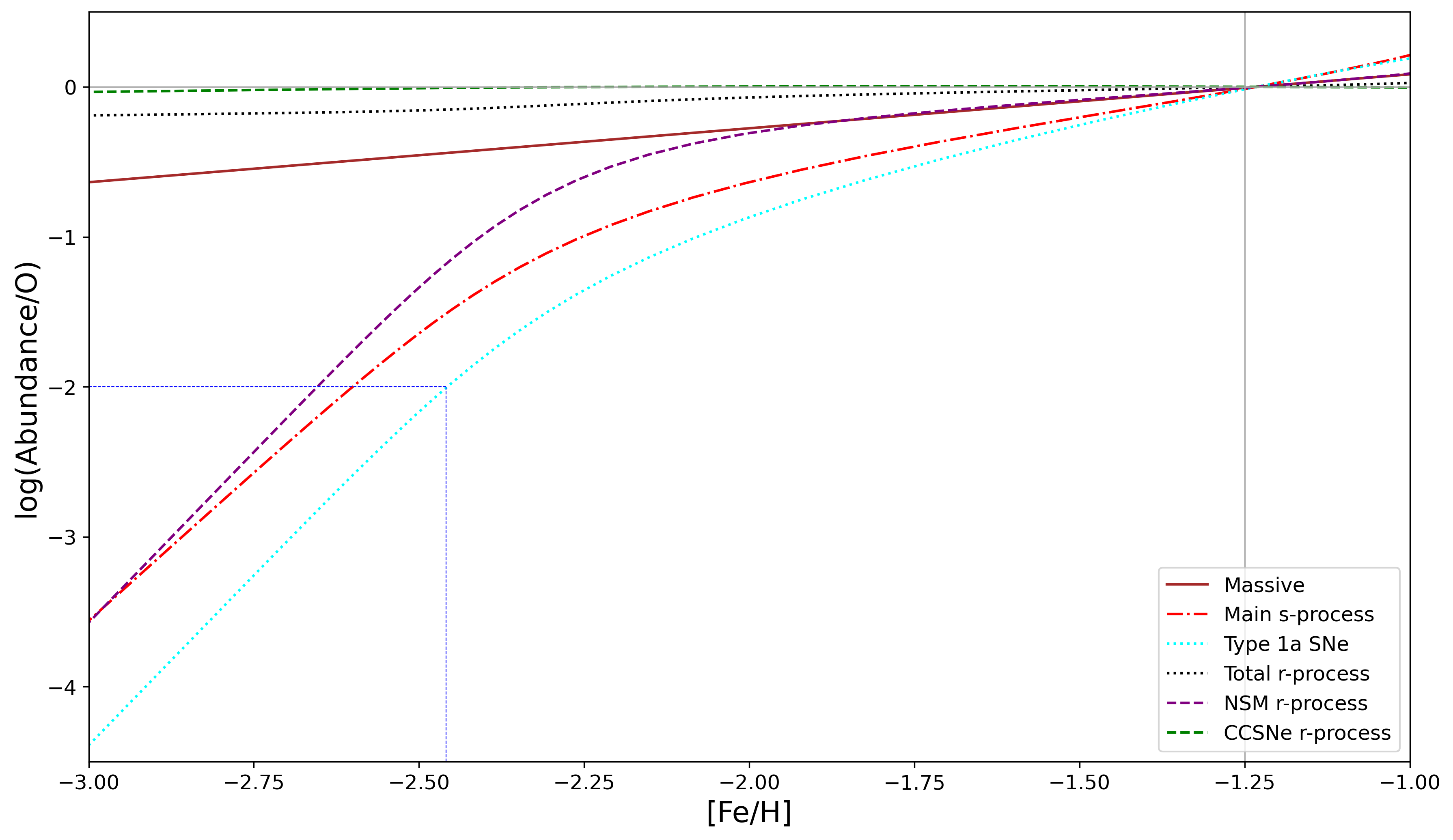}
\caption{The scaling functions of the model contributions relative to oxygen as functions of metallicity. The “Massive” line shows the scaling of the massive category’s contribution to $^{56}$Fe, and is normalized to the LSE contribution from this category only (as in \citealp{WH13}). The total \emph{r}-process is shown in addition to its modeled contributions from neutron star mergers and core-collapse supernovae. The Type \RNum{1}a onset metallicity, [Fe/H] = $-2.46$, is identified by where the Type \RNum{1}a scaling function drops to $1$ per cent its initial value (blue dotted line). \label{fig:Processes}}
\centering
\end{figure*}

The complete elemental scaling model is given in Figure \ref{fig:Scalings}. Whereas the model only considers s-process contributions to the s-only isotopes, s-process peaks for Z $\approx$ 38, Z $\approx$ 56, and Z $\approx$ 82 (corresponding to Sr, Ba, and Pb, respectively) can be seen. The odd-even effect can be seen below Z $\approx$ 10, but this trend becomes less distinct at higher atomic numbers (10 $\leq$ Z $\leq$ 30) compared to the MW (see Figure 9 of \cite{WH13}). Furthermore, the iron-peak seems to be less defined than the peak observed in Figure 9 of \cite{WH13}. These two differences in these trends occur due to the Type \RNum{1}a progenitor used in this model, as theoretical yields from sub-Chandrasekhar models tend to have more $\alpha$ element and lower iron-peak abundances compared to deflagration models (see, for e.g., \citealp{nomoto1997}, and Figure 6 of \citealp{kirby2019}). The scalings for H/He isotopes are approximate, since the LSE was computed using estimated parameters which are reasonably chosen yet not perfectly known. Given the gas exchange between dSphs and their environments, H and He may inherit larger uncertainties from the chose parameterizations in OMEGA. Figure \ref{fig:Processes} shows all the unique scaling functions against O considered in this work using the best-fitting free parameters determined in Section \ref{Fitting Scaling Model to Observational Data}. 

The best-fitting free parameters offer useful comparisons between the chemical evolution of the Sculptor dSph galaxy and the MW. As mentioned in Section \ref{Introduction}, iron production begins to dominate at an observed metallicity of [Fe/H] $\approx-1$ in the MW. The value of the free parameter $b$, the shifting factor for the Type \RNum{1}a tanh function, is less for Sculptor ($b$ = 1.073) than for the MW ($b$ = 2.722, \citealp{WH13}). This suggests that the Type \RNum{1}a onset occurs at a lower metallicity in dSphs than the MW and supports the conclusions of past chemical evolution models. We define the onset to occur when the Type \RNum{1}a abundance drops to 1 per cent of its LSE value  (in congruence with \citealp{WH13}). We find the onset to occur at [Fe/H] = $-$2.46, which is at a lower metallicity than the range $-$1.5 $<$ [Fe/H] $<$ $-$2.0 suggested by previous studies (eg. \citealp{fenner2003} and \citealp{reichert2020}). The onset value occurs at a lower metallicity than the observed metallicity of [Fe/H] $\approx-1$ in the MW \citep{smecker-hane1992, venn2012, kobayashi2011, kobayashi2020}, as well as the calculated value of [Fe/H] = $-$1.1 in \cite{WH13}.

Our model finds that Type \RNum{1}a SNe contribute more to the chemical evolution of iron peak isotopes in dSph galaxies than the MW compared to massive star contributions, in agreement with \cite{lanfranchi2003}, \cite{vincenzo2014}, and \cite{kirby2019}. In particular, we find the fraction of LSE Fe from Type \RNum{1}a SNe to be 0.86, which is greater than the MW (f = 0.69, \citealp{WH13}). This is close to the value from \cite{kirby2019} who found f$_{\text{\RNum{1}a}}$ $\approx$ 0.8 during Sculptor's LSE at [Fe/H] = $-$1.25 (see Figure 1 of \citealp{kirby2019}).

The decrease of the [$\alpha$/Fe] ratio after Type \RNum{1}a onset has been predicted to be steeper in dSphs than in the MW \citep{recchi2001, matteucci2008, vincenzo2014, theler2019}. The onset slope in this model is contained in the parameter $a$, the scaling factor for the Type \RNum{1}a function. We find a fit value ($a$ = 1.628) that is less than this parameter's value for the MW ($a$ = 5.024, \citealp{WH13}). This suggests the impact of Type \RNum{1}a SNe in driving the [Mg/Fe] ratio down to its LSE value occurred over a shorter metallicity range in Sculptor. There is dispute over the existence of [$\alpha$/Fe] low metallicity plateaus, however, so this conclusion should be handled with care. \cite{kirby2011} analyzed [$\alpha$/Fe] knees for numerous dSphs and only found significant knees for the [Ca/Fe] ratio in Sculptor and Ursa Minor but not for other $\alpha$-element ratios, such as [Mg/Fe] or [Si/Fe]. An absence of low metallicity plateaus in the [$\alpha$/Fe] ratios would suggest that Type \RNum{1}a SNe were dominant mechanisms throughout the entirety of the SFHs in dSphs \citep{kirby2011}. Clarification on this issue requires additional low metallicity data ([Fe/H] $< -2.5$). Present data, though sparse, suggests a [Mg/Fe] average that indeed plateaus, yet the metallicity which demarcates the [$\alpha$/Fe] knee is also  where the transition from sparse to more plentiful data occurs, which is concerning. 

Our results for the high metallicity behaviour of [Mg/Fe] are similar with \cite{kirby2011}. They observed that the slope of [Mg/Fe] at [Fe/H] $>$ $-$1 is less than the slope at [Fe/H] $<$ $-$1.5, which suggests some flattening of these ratios at higher metallicities. Our model shows that the [Mg/Fe] slope decreases beyond [Fe/H] = $-$1.25.

\cite{kirby2019} modelled the Type \RNum{1}a onset ([Fe/H]$_{\text{\RNum{1}a}}$) for various dSphs and determined a constant value of the ratio of an element relative to Fe before the onset ([X/Fe]$_{\text{CC}}$) using the Markov Chain Monte Carlo method. They fixed the initial value of [Fe/H]$_{\text{\RNum{1}a}}$ to $-$2 and calculated an angle and perpendicular offset for each element X using a linear fit for [X/Fe] vs [Fe/H]. For Sculptor, they calculated [Fe/H]$_{\text{\RNum{1}a}}$ = $-$2.12, which is 0.4 dex above our calculated onset value of [Fe/H]$_{\text{\RNum{1}a}}$ = $-$2.5 relative to the solar abundance decomposition. They also calculated [Mg/Fe]$_{\text{CC}}$ to be +0.56 $\pm$ 0.01, which is similar with our model's [Mg/Fe]$_{\text{CC}}$ value of approximately 0.5 determined by OMEGA (Figure \ref{fig:Mg}). Our model's [Mg/Fe]$_{\text{CC}}$ value of 0.5 is also less than the [Mg/Fe]$_{\text{CC}}$ value for the MW which plateaus at approximately 0.57 \citep{WH13}. This supports the idea that a system with a low [Mg/Fe]$_{\text{CC}}$ value at low metallicities undergoes a slow, extended SF \citep{fenner2006}, since Sculptor's SFH took place for an extended period of time from 14 gyr to 7 gyr ago during a single formation event and gradually decreased afterwards \citep{deboer2012a}. \cite{kirby2011} also observed that the average value of [$\alpha$/Fe] drops from +0.4 at [Fe/H] = $-$2.5 to 0.0 at [Fe/H] = $-$1.2. Our model also shows this trend, but over predicts the [$\alpha$/Fe] value by about 1 dex at low metallicities. Similarly, \cite{mashonkina2017} noted that dSphs and the MW exhibit a plateau of [$\alpha$/Fe] $\approx$ 0.3 at [Fe/H] $\leq$ $-$2.5, which is 2 dex below our model's plateau at the same metallicity range. 

Figure \ref{fig:Mg} shows that the best-fitting model curve is at the edge of the parameter space (shown as the cyan hue in the background). This is likely because the Mg LSE value is below the data average by $\approx$ 0.15 dex, which results in a parameter space that is slightly below the data average from $-$2 $\leq$ [Fe/H] $\leq$ $-$1.25. Since the parameters $a$ and $b$ most greatly influence the Mg model at this region, these parameters were shifted to the top of the parameter space by the machine learning algorithm to best match the data average. It is possible that our previous comparisons between the best-fit values of $a$ and $b$ for Sculptor and the MW may change significantly if we had a Mg LSE value that more closely matched the data average. To test for such an effect, we recomputed the best-fit values for $a$ and $b$ assuming the LSE model gives the observed [Mg/Fe] value at [Fe/H] = $-1.25$ and found that $a$ = 3.059 and $b$ = 0.898 (compared to $a = 1.628$ and $b = 1.073$). The higher value of $a$ suggests a shallower slope for Type \RNum{1}a, and the lower value of $b$ suggests a lower metallicity for Type \RNum{1}a onset. Nevertheless, both sets of values are lower than the corresponding MW values, hence the previous comparisons between the test for Sculptor and the MW models hold.

The model predicts that [Ba/Fe] $\approx$ 0.1 at [Fe/H] = $-$1, which is above the MW value by 0.4 dex. This supports the conclusions of previous studies that AGB stars are the most likely sites for \emph{s}-process production in Sculptor due to the large [Ba/Fe] ratios at high metallicities \citep{fenner2006, tolstoy2009, cohen2009, venn2012, skuladottir2019, Skuladottir2020, theler2019, reichert2020}. Whereas the model for Ba from \cite{WH13} has a value of [Ba/Fe] $\approx$ 0.2 at [Fe/H] = $-$1, which somewhat contradicts this claim, their model slightly decreases after [Fe/H] $\geq$ $-$1, while our model shows no such trend. Thus, our model does indeed predict higher [Ba/Fe] values, only after [Fe/H] = $-$1. 

The transition from \emph{r}-process to \emph{s}-process dominated contributions can be identified by the intersection between the \emph{s}-process and \emph{r}-process model lines. We find the intersection for [Ba/Fe] to occur at [Fe/H] $\approx$ $-$2.22, which is a bit lower than the value [Fe/H] = $-$2 suggested by previous studies \citep{tolstoy2009, venn2012, skuladottir2019} and [Fe/H] = $-$1.57 suggested by \cite{reichert2020}. The difference likely arises due to the different numerical methodologies in our work compared or different definitions of the \emph{r}-process to \emph{s}-process transitions. For example, \cite{reichert2020} performed a least squares fit on [Ba/H] to [Fe/H] and found that [Eu/Ba] decreased at a metallicity of $-$0.57 for Sagittarius, $-$1.04 for Fornax, and $-$ 1.57 for Sculptor. This suggests that more massive galaxies may have an \emph{s}-process onset that occurs later than less massive galaxies. For the MW, \cite{WH13} found this transition to occur at [Fe/H] $\approx$ $-$1.9, which is consistent with \cite{reichert2020} and \cite{cohen2009} who suggested that the MW should have an \emph{s}-process transition that is later than that of dSphs. 

Previous studies have also suggested that AGB stars began dominating the chemical evolution of dSphs on the same timescales as Type \RNum{1}a SNe. Our model finds a similar trend; the \emph{s}-process to \emph{r}-process transition ([Fe/H] = $-$2.22) occurs near (though above) the Type \RNum{1}a onset ([Fe/H] = $-$2.46) relative to the solar abundance pattern \citep{Lodders2020}. They also seem to be co-produced (to within 0.2 dex) after [Fe/H] $\approx$ $-$1.75 from Figure \ref{fig:Processes}. Another indication that AGB stars and Type \RNum{1}a SNe dominate at similar timescales is a flat [Ba/Fe] ratio at high metallicities \citep{venn2012, Skuladottir2020}. The Ba model shows a flat ratio between $-2 < $ [Fe/H] $< -1$ and increases after [Fe/H] $>$ $-$1, suggesting that AGB enrichment dominates over Type \RNum{1}a SNe enrichment in Sculptor at high metallicities (see also Figure \ref{fig:Ba}), though care should be taken since high metallicity data is sparse. We also find that the main \emph{s}-process in AGB stars have a somewhat stronger metallicity dependence in Sculptor than the MW, as the best-fitting $h$-value of Sculptor ($h$ = 1.798) is greater than the best-fitting $h$-value of the MW ($h$ = 1.509, \citealp{WH13}).

The [Eu/Fe] model ratio is above zero at low metallicities and seems to plateau at a value of [Eu/Fe] $\approx$ 0.7. This is similar to the value of [Eu/Fe] $\approx$ 0.8 for the MW \citep{WH13}, which supports previous studies suggesting similar trends in both galaxies \citep{Skuladottir2020}. The [Eu/Fe] ratio begins to decline at [Fe/H] $\gtrsim$ $-$2.5, due to the Type \RNum{1}a onset. 

Surprisingly, the primary process exponent parameter, $p$, was found to be much lower ($p$ = 0.391) than the MW value ($p$ = 0.938, \citealp{WH13}). A lower $p$-value corresponds to a larger decrease in the [Eu/Fe] ratio as [Fe/H] increases beyond Type \RNum{1}a onset. The value of the parameter $g$, the fraction that NSM contributes to \emph{r}-process isotopes, suggests that NSM only contributes about 30 per cent to the LSE of Eu, which agrees with previous studies that propose CCSNe, rather than NSM, are the dominant \emph{r}-process progenitors in dSphs \citep{tolstoy2009, duggan2018, skuladottir2019, reichert2020, Skuladottir2020}. 

\begin{figure*}
\includegraphics[width=6.5in]{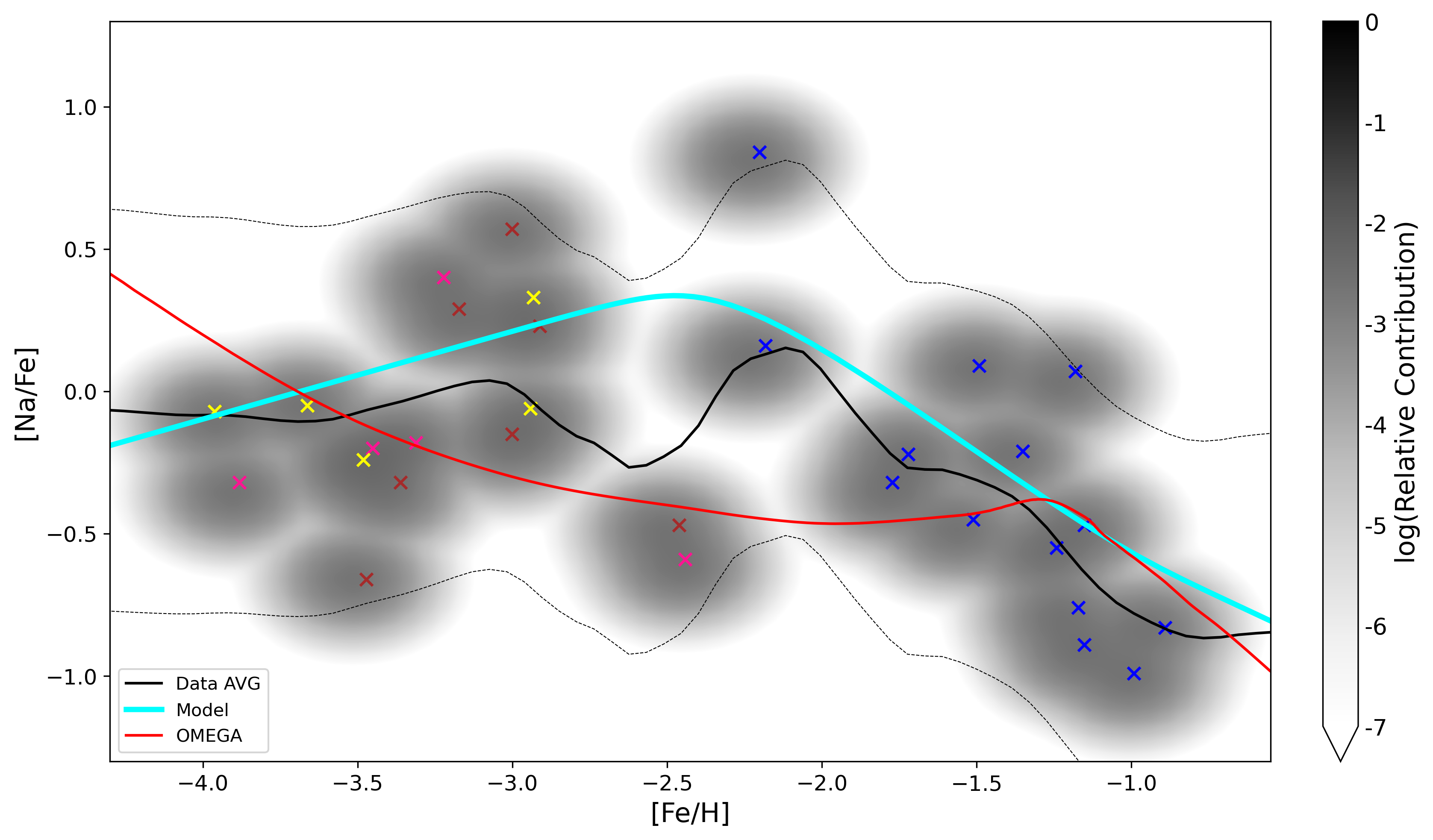}
\caption{The resulting model for [Na/Fe] determined after fitting parameters. The figure follows the same convention as Figure \ref{fig:Mg}. Plotted data is from \citealp{tafelmeyer2010} (yellow), \citealp{starkenburg2013} (brown), \citealp{jablonka2015} (pink), and \citealp{hill2019} (blue).}\label{fig:Na}
\centering
\end{figure*}

\begin{figure*}
\includegraphics[width=6.5in]{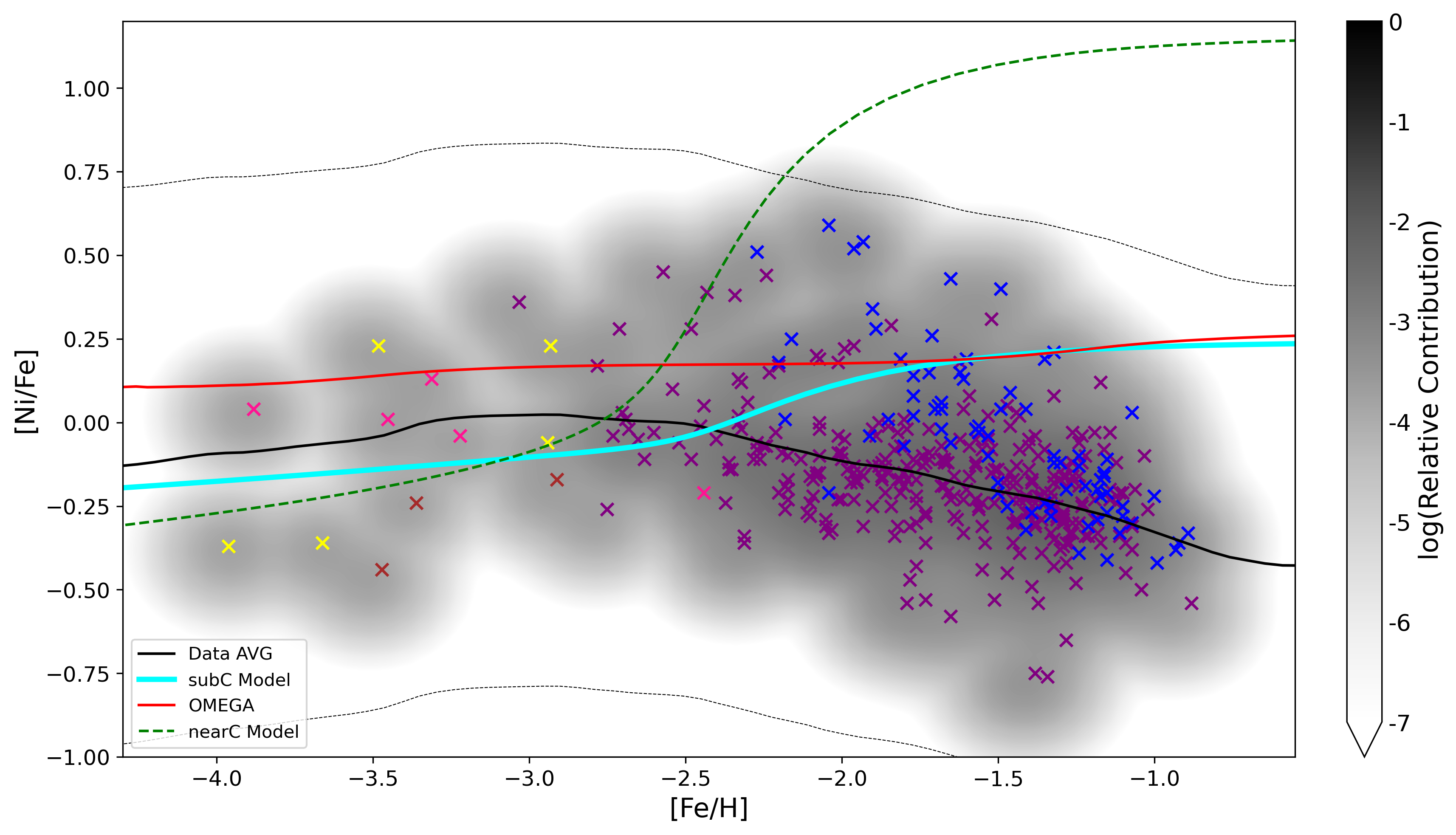}
\caption{The resulting models for [Ni/Fe] determined after fitting parameters. Both sub-Chandrasekhar (subC) and near-Chandrasekhar (nearC) model curves are shown. The figure follows the same convention as Figure \ref{fig:Mg}. Plotted data is from \citealp{starkenburg2013} (brown), \citealp{jablonka2015} (pink), \citealp{tafelmeyer2010} (yellow), \citealp{kirby2018} (purple), and \citealp{hill2019} (blue).
 \label{fig:Ni}}
\centering
\end{figure*}

Figure \ref{fig:Na} and Figure \ref{fig:Ni} show the model for [Na/Fe] and [Ni/Fe], respectively. Our model is in reasonable agreement to observed Na trends, with [Na/Fe] increasing at low metallicities ([Fe/H] $<$ $-$2.5) and decreasing at higher metallicities as suggested by \cite{cohen2009} and  \cite{tolstoy2009}. Pauce Na data, however, makes a detailed comparison difficult. Whereas the model for Ni is similar to observed trends at low metallicities ([Fe/H] $\leq$ $-$3), it appears overproduced at higher metallicities ([Fe/H] $>$ $-$2). Both Na and Ni models seem to perform at least as well as OMEGA in representing the data averages.

\begin{figure}
\includegraphics[width=\columnwidth]{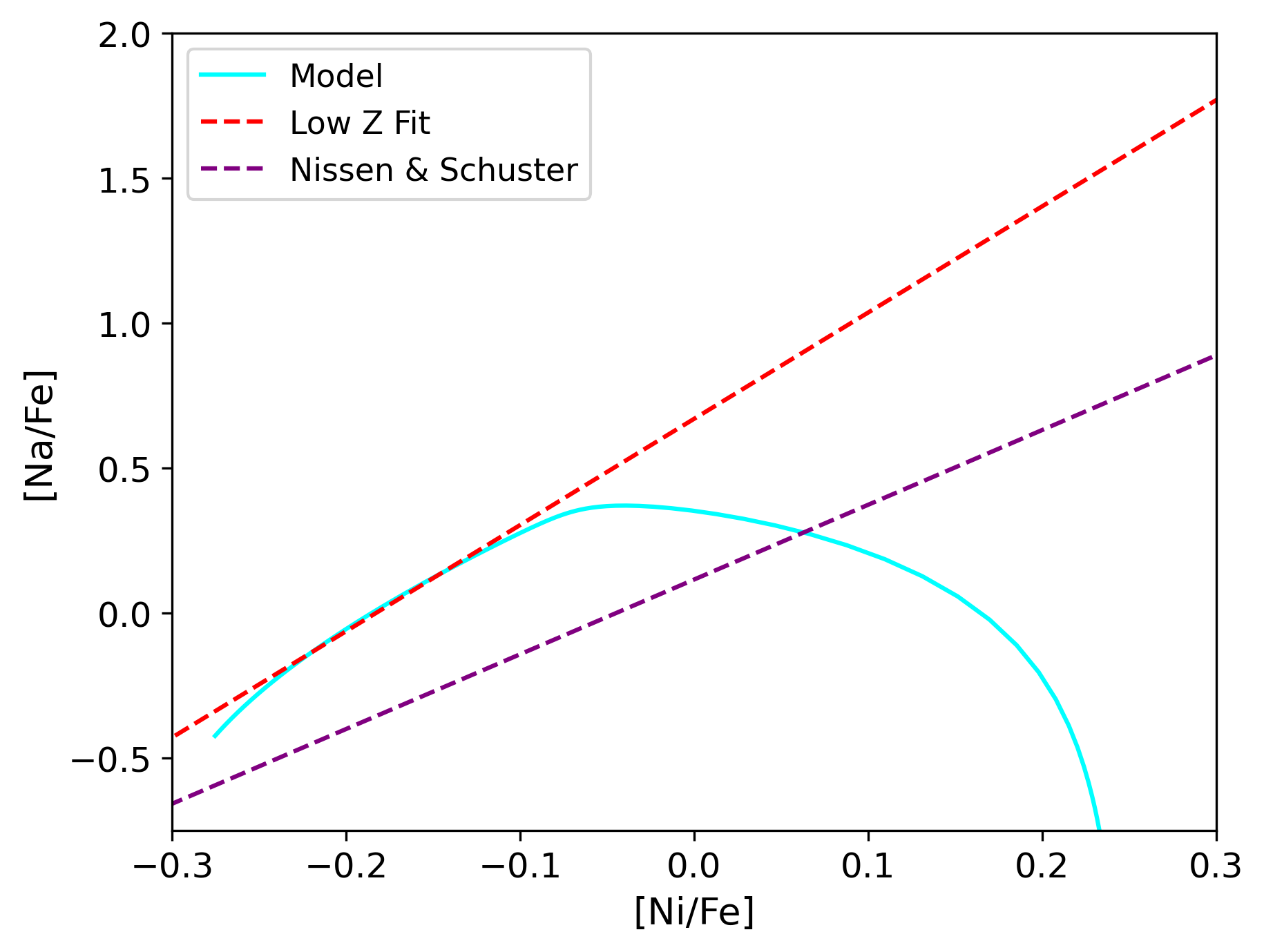}
\caption{The best-fitting line for the Na-Ni correlation in Sculptor compared to the the best-fitting line in the MW, determined by \citealp{nissen1997}. The Sculptor best-fitting line was constrained to [Fe/H] $\leq$ $-$2.5 ($-$0.23 $\leq$ [Ni/Fe] $\leq$ $-$0.075). \label{fig:NaNi}}
\centering
\end{figure}

\cite{nissen1997} noticed a Na/Ni correlation for MW halo stars, and this correlation was also suggested to occur in dSphs by \cite{tolstoy2009}. Figure \ref{fig:NaNi} shows this correlation in Sculptor. Whereas \cite{nissen1997} determined the best-fitting line to be [Na/Fe] = 2.577$\cdot$[Ni/Fe] + 0.116 for the MW, we find the best-fitting line to be [Na/Fe] = 3.663$\cdot$[Ni/Fe] + 0.670 during the early evolution of Sculptor ($-$0.23 $\leq$ [Ni/Fe] $\leq$ $-$0.075) with a $\chi^2$ = 0.002. We constrain this fit to [Fe/H] $\leq$ $-$2.5 because after this point, the model overproduces [Ni/Fe], as seen in the sharp downward trend in Figure \ref{fig:NaNi} and sharp upward trend in Figure \ref{fig:Ni}. The larger slope for our best-fitting line suggests that at low metallicities, Na and Ni were co-produced but Na was produced at a higher rate in Sculptor than in the MW.

The overproduction of Ni at higher metallicities shown in Figure \ref{fig:Ni} is likely due to the discrepancy between the observed [Ni/Fe] data average and the LSE [Ni/Fe] value from OMEGA at [Fe/H] = $-$1.25, a difference of $\approx$ 0.5 dex. We find [Ni/Fe] $\approx$ 0.2 at [Fe/H] = $-$1.5, which is consistent with \cite{kirby2019} who found that the sub-Chandrasekhar yields by \cite{leung2020} resulted in [Ni/Fe] $\approx$ 0.1 at [Fe/H] = $-$1.5. To see if a different Type \RNum{1}a progenitor improved our model for Ni, we also tested a near-Chandrasekhar model with a deflagration-detonation transition \citep{leung2018}. Using this Type \RNum{1}a progenitor, the [Ni/Fe] was \emph{more} overproduced ([Ni/Fe] $\approx$ 1) at [Fe/H] = $-$1.5. Although neither of these Type \RNum{1}a progenitors was able to reproduce observed [Ni/Fe] trends at higher metallicities, the sub-Chandrasekhar yields produced better results which agrees with \cite{kirby2019}. 


The model for [C/Fe] is vastly overproduced at all metallicities. [Sc/Fe] fits well with observed data, similar to [Na/Fe]. The model for [O/Fe] has a similar shape with observed O abundances, but overproduces them by about 0.4 dex from $-$3 $<$ [Fe/H] $<$ $-$1. [Si/Fe] shows similar trends, but is overproduced by about 0.3 dex in the same metallicity range. 

The model for iron-peak elements can also be evaluated. Co is the best fitting iron-peak element compared to observed data throughout all metallicities. [Co/Fe] slightly decreases at [Fe/H] $>$ $-$2.5, which was noted previously by \cite{cohen2009, cohen2010, kirby2019}. This behaviour supports the idea that Type \RNum{1}a SNe are dominant in dSphs at higher metallicities, which is why Fe production begins dominating the production of other iron-peak elements \citep{kirby2019}. The model overproduces [Cr/Fe] by about 0.75 dex at low metallicities but is only slightly overproduced at higher metallicities by about 0.2 dex ([Fe/H] $>$ $-$2.5). Both [Cr/Fe] and [Co/Fe] seem to plateau at a value of 0 at higher metallicities, suggesting that Type \RNum{1}a SNe produce Fe and other iron-peak elements in roughly equal amounts. The model for [Mn/Fe] slopes downward at all metallicities, which is the opposite trend than is observed. However, Mn data is sparse and is only available at low metallicities ($-$4 $\leq$ [Fe/H] $\leq$ $-$2.25). \cite{north2012} also predicted Mn to be subsolar. The model for [Zn/Fe] is similar to Mn; it decreases at all metallicities, as suggested by \cite{cohen2009, skuladottir2017, kirby2019}. 

\begin{figure}
\includegraphics[width=\columnwidth]{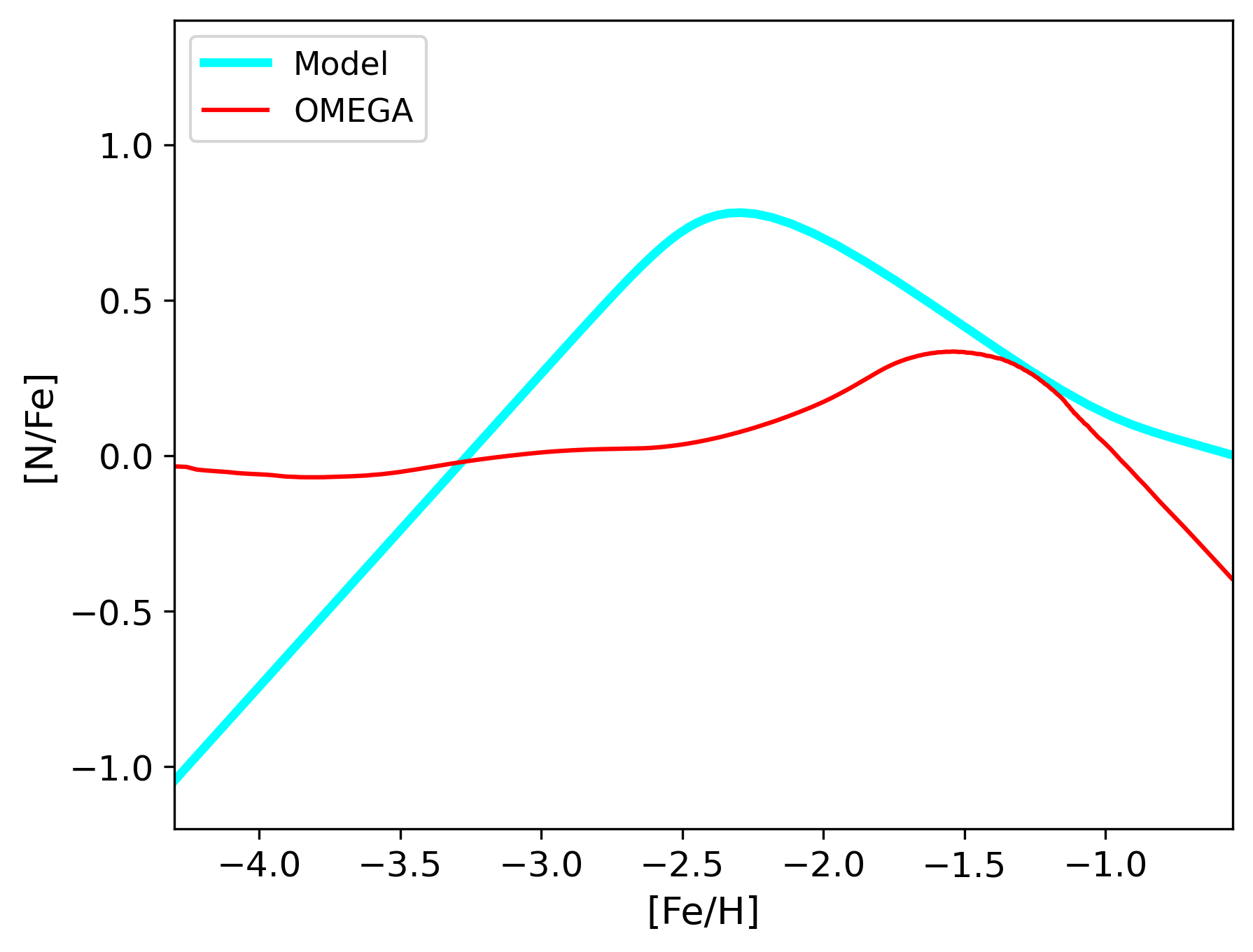}
\caption{The resulting model for [N/Fe] using the best-fitting parameters found in Section \ref{Fitting Scaling Model to Observational Data}. The model line is shown by the solid cyan line, and the OMEGA GCE model is shown by the solid red line. \label{fig:N}}
\centering
\end{figure}

It is useful to compare the effects of using rotating and non-rotating yields on the [N/Fe] model at the intermediary point [Fe/H] = $-$3. Figure \ref{fig:N} shows the model and the OMEGA LSEM for N. Below [Fe/H] $< -3.3$, OMEGA produces more N. Since the model's intermediary point uses non-rotating yields and OMEGA uses rotating yields, this trend is consistent with previous studies that suggest that massive stars with faster rotation produce more N \citep{chiappini2006}. These studies have also shown that rotating yields for N are strongly dependent on metallicity below [Z] $\leq$ $-$5, though our model is not constrained by data at such low metallicities and hence does not seek to describe an average chemical history there (if indeed an average is even well-defined; see \citealp{WH13} for further discussion). At [Fe/H] = $-$3, the model gives [N/Fe] = 0.25 and OMEGA gives [N/Fe] = 0.0. These values suggest that at the intermediary point, the choice of using non-rotating yields for the model moderately impacts the production of N, although this choice significantly impacts the production of other elements such as Na, Mg, and Ni (see Figure \ref{fig:Rotation_minus3}). Figure \ref{fig:N} shows that below [Fe/H] = $-2.5$, the model line is sloped due to the suggested secondary nature of $^{14}$N for non-rotating yields, and the OMEGA curve is constant due to the suggested primary nature for rotating yields \citep{prantzos2018}. Whereas there is no N data to determine which curve is more accurate, the discrepancies between the two curves due to rotating and non-rotating yields suggests that the N model may not be a good representation of the evolution of N in Sculptor. OMEGA is likely more accurate in the low metallicity regime, where rotating stars should increase the production of N \citep{chiappini2006}. We choose to use non-rotating yields from \cite{heger2010} for the intermediary fixed point, however, because it better models Na, Mg, and Ni (see Figure \ref{fig:Rotation_minus3}), and we accept that the model may not be accurate for N.

Finally, we compare our model to a traditional GCE model (see Figures \ref{fig:Mg}, \ref{fig:Eu}, \ref{fig:Ba}, \ref{fig:Na}, \ref{fig:Ni}). The GCE model used for comparison is the one also used to provide the LSE abundance for Sculptor, hence our model values agree with OMEGA's at [Fe/H] $= -1.25$. Our Mg/Fe model tracks the data average more consistently than OMEGA. Our Ba/Fe model and OMEGA both track the observed Ba trends, though OMEGA likely underproduces the ratio at higher metallicities ([Fe/H] $>$ $-$1.25), although the data is sparse. OMEGA does not fit the Eu/Fe ratio at low metallicities, whereas our model performs significantly better in this regime. Our model for Na/Fe roughly captures the observed average of Na, though data is quite sparse. OMEGA shows a similar trend to our model at higher metallicities ([Fe/H] $>$ $-$1.25), and it is difficult to assess from available data which performs better. If Na is dominated by massive stars, then one would expect the ratio against Fe to be above zero. It is not clear why a [Na/Fe] ratio would be a decreasing function prior to Type \RNum{1}a onset. Both our model for Ni/Fe and OMEGA fail to reproduce observed Ni trends, as both overestimate [Ni/Fe] at higher metallicities ([Fe/H] $>$ $-$2.5). Finally, even though there is no N data to determine whether our N/Fe model or OMEGA is more accurate, it is expected that OMEGA performs better, especially in the low-metallicity regime ([Fe/H] $\leq$ $-2.5$). This is due to our model using non-rotating yields for the intermediary fixed point which likely causes an underproduction of N.

\begin{table}
    \centering
    \caption{\label{tab:Parameters}Optimized Parameter Values}
    \begin{tabular}{*3l}
        \toprule
        Parameter & Best-fitting Value  & Description \\
        \midrule
        $a$ & 1.628 & Type \RNum{1}a tanh scaling factor \\
        $b$ & 1.073 & Type \RNum{1}a tanh shifting factor       \\
        $f$ & 0.863 & Fraction of $^{56}$Fe$_{\text{LSE}}$ from Type \RNum{1}a \\
        $p$ & 0.391 & Primary process exponent       \\
        $h$ & 1.798 & \emph{hs}-process exponent \\
        $c$ & 5.343 & NSM tanh scaling factor \\
        $d$ & 0.450 & NSM tanh shifting factor \\
        $g$ & 0.303 & Fraction of Eu$_{\text{LSE}}$ from NSM \\
        \bottomrule
    \end{tabular}
\end{table}

\section{Conclusions}
\label{Conclusions}
A model has been constructed that describes the average isotopic decomposition of the Sculptor dwarf spheroidal galaxy for contributions from massive stars, Type \RNum{1}a SNe, main \emph{s}-process peak, and the \emph{r}-process, and the procedure follows the methodology described in \cite{WH13}. Astrophysical processes that dictate the chemical evolution of galaxies were discussed and compared between the MW and dSphs. Parametric equations were assigned to these processes and scaled between BBN and the Sculptor LSE decomposition, which was computed using OMEGA at [Fe/H] = $-$1.25 \citep{cote2017}. The isotopic scalings were summed to elemental scalings and compared with Sculptor data to fix the free parameters of the model. The result is a complete isotopic abundance decomposition of Sculptor at any desired metallicity for the processes considered. 

While many assumptions and approximations were made when constructing the model, this work still offers many comparisons between dSphs and spiral galaxies, with Sculptor and the MW as candidates: 
\begin{itemize}
  \item The Type \RNum{1}a onset occurs at a lower metallicity in Sculptor than the MW. Numerically, we define the onset to occur when the Mg abundance from Type \RNum{1}a SNe is 1 per cent of its Type \RNum{1}a late-stage evolution value. This occurs at [Fe/H] = $-$2.46, which is less than the observed Type \RNum{1}a onset of the MW ([Fe/H] $\approx$ $-$1). 
  \item The model predicts that Type \RNum{1}a SNe contribute $\approx$ 86 per cent to the $^{56}$Fe LSE abundance, which is greater than predicted values of $\approx$ 70 per cent in the MW \citep{WH13}. Thus, Type \RNum{1}a SNe contribute more to the chemical evolution of dSphs compared to the MW, in agreement with past chemical evolution models.
  \item The slope of the [$\alpha$/Fe] ratio after the Type \RNum{1}a onset is steeper in Sculptor than the MW, suggesting that the impact of Type \RNum{1}a SNe occurred over a shorter metallicity range. 
  \item The model predicts that sub-Chandrasekhar Type \RNum{1}a SNe are the dominant Type \RNum{1}a progenitor in Sculptor, in agreement with \cite{kirby2019}.
   \item The \emph{r}-process to \emph{s}-process transition occurs at [Fe/H] $\approx$ $-$2.22, which disagrees with previous studies that suggest ranges from $-$1.57 $<$ [Fe/H] $<$ $-$2.0. This metallicity is less than the MW transition of [Fe/H] $\approx$ $-$1.9 \citep{WH13}, which is consistent with  the idea that a more massive galaxy should have a later transition. 
  \item The model suggests that AGB stars are more dominant in Sculptor compared to the MW and are the most likely sites for \emph{s}-process production. AGB enrichment also seems to occur at similar timescales as Type \RNum{1}a SNe since the \emph{r}-process to \emph{s}-process transition ([Fe/H] = $-$2.22) occurs at the same time as the Type \RNum{1}a onset ([Fe/H] = $-$2.5), as suggested by previous studies. However, a [Ba/Fe] plateau is not observed at high metallicities since AGB contribution is more pronounced than Type \RNum{1}a SNe at this range.
  \item The \emph{r}-process evolution in Sculptor appears to be similar to the evolution in the MW, with similar [Eu/Fe] plateaus at low metallicities and a [Eu/Fe] decline beginning at Type \RNum{1}a onset. 
  \item The model shows that NSM only contribute $\approx$ 30 per cent to Eu$_{\text{LSE}}$, supporting previous studies that CCSNe, rather than NSM, are the most dominant progenitor for \emph{r}-process production in Sculptor.
  \item The model tracks the data average more closely than OMEGA for Mg, Ba, Eu, and Na, and performs equally well in reproducing observed Ni trends. It is likely that the model for N is not as accurate, since non-rotating yields were used for the intermediary fixed point at [Fe/H] = $-$3.
\end{itemize}

In addition to offering comparisons between the chemical evolution of dSphs and the MW, this model also provides isotopic inputs for future nucleosynthesis studies of dSphs. Future works can explore this model for other dSphs to provide further insight into the chemical evolution differences between dSphs and spiral galaxies and ultimately further our understanding of hierarchical galactic formation. 

\section*{Data Availability}
The data that support the findings of this study are available in the public domain. Table \ref{tab:Data} lists the main data sources used in this work. Additional results from this work (e.g., elemental model plots) are available from the corresponding author upon request.	

\begin{table*}
    \centering
    \captionsetup{justification=centering}
    \caption{\label{tab:Data}Third-Party Data Sources}
    \begin{tabular}{*3l}
        \toprule
        Description & Publication  & Data Link \\
        \midrule
        Abundances of $\alpha$ elements for $\approx$ 400 stars & \cite{kirby2009} & \href{https://vizier.u-strasbg.fr/viz-bin/VizieR-3?-source=J/ApJ/705/328&-out.max=50&-out.form=HTML$\%$20Table&-out.add=_r&-out.add=_RAJ,_DEJ&-sort=_r&-oc.form=sexa}{Link} \\
        Abundances of iron-peak, neutron-capture, and $\alpha$ elements for two stars & \cite{tafelmeyer2010} & \href{https://www.aanda.org/articles/aa/full_html/2010/16/aa14733-10/aa14733-10.html}{Link} (Table 7)\\
        Abundances of iron-peak and $\alpha$ elements for seven stars & \cite{starkenburg2013} & \href{https://www.aanda.org/articles/aa/full_html/2013/01/aa20349-12/aa20349-12.html}{Link} (Table 6) \\
        Abundances of iron-peak, neutron-capture, and $\alpha$ elements for 99 stars & \cite{hill2019} & \href{https://vizier.u-strasbg.fr/viz-bin/VizieR-3?-source=J/A\%2bA/626/A15/tablec5&-out.max=50&-out.form=HTML\%20Table&-out.add=_r&-out.add=_RAJ,_DEJ&-sort=_r&-oc.form=sexa}{Link} \\
        Abundances of Zn for 100 stars & \cite{skuladottir2017} & \href{https://vizier.u-strasbg.fr/viz-bin/VizieR-3?-source=J/A\%2bA/606/A71/table2&-out.max=50&-out.form=HTML\%20Table&-out.add=_r&-out.add=_RAJ,_DEJ&-sort=_r&-oc.form=sexa}{Link} \\
        Abundances of iron-peak elements for $\approx$ 300 stars & \cite{kirby2018} & \href{https://vizier.u-strasbg.fr/viz-bin/VizieR-3?-source=J/ApJS/237/18/table3&-out.max=50&-out.form=HTML\%20Table&-out.add=_r&-out.add=_RAJ,_DEJ&-sort=_r&-oc.form=sexa}{Link} \\
        Abundances of Ba for 120 stars & \cite{duggan2018} & \href{https://vizier.u-strasbg.fr/viz-bin/VizieR-3?-source=J/ApJ/869/50/table6&-out.max=50&-out.form=HTML\%20Table&-out.add=_r&-out.add=_RAJ,_DEJ&-sort=_r&-oc.form=sexa}{Link} \\
        Abundances of neutron-capture elements for 98 stars & \cite{skuladottir2019} & \href{https://vizier.u-strasbg.fr/viz-bin/VizieR-3?-source=J/A\%2bA/631/A171/tableb1&-out.max=50&-out.form=HTML\%20Table&-out.add=_r&-out.add=_RAJ,_DEJ&-sort=_r&-oc.form=sexa}{Link} \\
        Abundances of iron-peak, neutron-capture, and $\alpha$ elements for 5 metal-poor stars & \cite{jablonka2015} & \href{https://www.aanda.org/articles/aa/pdf/2015/11/aa25661-15.pdf}{Link}  (Table 7) \\
        OMEGA GCE code & \cite{cote2017} & \href{http://nugrid.github.io/NuPyCEE/SPHINX/build/html/omega.html#}{Link} \\
        Massive star theoretical yields & \cite{heger2010} & \href{https://2sn.org/starfit/}{Link} \\
        Type \RNum{1}a SNe theoretical yields (sub-Chandrasekhar) & \cite{leung2020} & \href{https://iopscience.iop.org/article/10.3847/1538-4357/ab5c1f}{Link} (Table 7)\\
        Type \RNum{1}a SNe theoretical yields (near-Chandrasekhar) & \cite{leung2018} & \href{https://iopscience.iop.org/article/10.3847/1538-4357/aac2df}{Link} (Table 4)\\
        AGB contributions & \cite{karakas2010} & \href{https://vizier.u-strasbg.fr/viz-bin/VizieR-3?-source=J/MNRAS/403/1413/yields&-out.max=50&-out.form=HTML\%20Table&-out.add=_r&-out.add=_RAJ,_DEJ&-sort=_r&-oc.form=sexa}{Link}\\
        CCSNe contributions & \cite{limongichieffi2018} & \href{https://vizier.u-strasbg.fr/viz-bin/VizieR-3?-source=J/ApJS/237/13/table9&-out.max=50&-out.form=HTML\%20Table&-out.add=_r&-out.add=_RAJ,_DEJ&-sort=_r&-oc.form=sexa}{Link}\\
        \bottomrule
    \end{tabular}
\end{table*}


\section{Acknowledgements}
We are grateful to the referee for several insightful comments and suggestions. We also thank Benoit C\^{o}t\'{e} and Amanda Karakas for their assistance with this work.



\bibliographystyle{mnras}
\bibliography{revised} 





\bsp	
\label{lastpage}
\end{document}